\documentclass[aip,jcp,reprint,showkeys,floatfix]{revtex4-2}

\usepackage{amssymb}
\usepackage{amsmath}
\usepackage{amsfonts}
\usepackage{booktabs}
\usepackage{enumerate}
\usepackage[hidelinks]{hyperref}
\usepackage{mathrsfs}
\usepackage{mathtools}
\usepackage{mhchem}
\usepackage{multirow}
\usepackage{physics}
\usepackage{placeins}
\usepackage{tensor}
\usepackage{graphicx}
\usepackage{siunitx}
\usepackage[capitalize]{cleveref}
\allowdisplaybreaks

\DeclareSIUnit{\angs}{\text{\AA}}
\DeclareSIUnit{\hartree}{\ensuremath{E_\mathrm{h}}}
\DeclareSIUnit{\millihartree}{\milli\hartree}
\DeclareSIUnit{\microhartree}{\micro\hartree}

\newcommand{\Pu}{P{\mathrel{\uparrow}}}
\newcommand{\Pd}{P{\mathrel{\downarrow}}}
\newcommand{\Qu}{Q{\mathrel{\uparrow}}}
\newcommand{\Qd}{Q{\mathrel{\downarrow}}}
\newcommand{\Ru}{R{\mathrel{\uparrow}}}
\newcommand{\Rd}{R{\mathrel{\downarrow}}}
\newcommand{\Su}{S{\mathrel{\uparrow}}}
\newcommand{\Sd}{S{\mathrel{\downarrow}}}

\newcommand{\sm}{supplementary material}
\newcommand{\latin}[1]{{\textit{#1}}}

\begin{document}

\title{Symmetry Dilemmas in Quantum Computing for Chemistry:\linebreak
A Comprehensive Analysis}

\author{Ilias Magoulas}
\email{ilias.magoulas@emory.edu}
\affiliation{Department of Chemistry and Cherry Emerson Center for Scientific 
Computation, Emory University, Atlanta, Georgia 30322, USA}

\author{Muhan Zhang}
\affiliation{Department of Chemistry and Cherry Emerson Center for Scientific 
Computation, Emory University, Atlanta, Georgia 30322, USA}

\author{Francesco A.\ Evangelista}
\affiliation{Department of Chemistry and Cherry Emerson Center for Scientific 
Computation, Emory University, Atlanta, Georgia 30322, USA}

\begin{abstract}
    Symmetry adaptation, universality, and gate efficiency are central but often competing requirements in quantum algorithms for electronic structure and many-body physics.
    For example, fully symmetry-adapted universal operator pools typically generate long and deep quantum circuits, gate-efficient universal operator pools generally break symmetries, and gate-efficient fully symmetry-adapted operator pools may not be universal.
    In this work, we analyze such symmetry dilemmas both theoretically and numerically.
    On the theory side, we prove that the popular, gate-efficient operator pool consisting of singlet spin-adapted singles and perfect-pairing doubles is not universal when spatial symmetry is enforced.
    To demonstrate the strengths and weaknesses of the three types of pools, we perform numerical simulations using an adaptive algorithm paired with operator pools that are (i) fully symmetry-adapted and universal, (ii) fully symmetry-adapted and non-universal, and (iii) breaking a single symmetry and are universal.
    Our numerical simulations encompass three physically relevant scenarios in which the target state is (i) the global ground state, (ii) the ground state crossed by a state differing in multiple symmetry properties, and (iii) the ground state crossed by a state differing in a single symmetry property.
    Our results show when symmetry-breaking but universal pools can be used safely, when enforcing at least one distinguishing symmetry suffices, and when a particular symmetry must be rigorously preserved to avoid variational collapse.
    Together, the formal and numerical analysis provides a practical guide for designing and benchmarking symmetry-adapted operator pools that balance universality, resource requirements, and robust state targeting in quantum simulations for chemistry.
\end{abstract}

\maketitle

\section{Introduction}

A foundational concept in quantum computing is that of a universal gate set,\cite{Deutsch.1989.10.1098/rspa.1989.0099,Deutsch.1995.10.1098/rspa.1995.0065,DiVincenzo.1995.10.1103/PhysRevA.51.1015,Barenco.1995.10.1103/PhysRevA.52.3457} \latin{i.e.}, a set of quantum gates whose finite compositions can approximate any unitary operation on any number of qubits to arbitrary precision.
Equivalently, a gate set is universal if, for any number of qubits $n$, a finite product of these gates (\latin{i.e.}, a circuit) can transform, up to a global phase, any normalized $n$-qubit state into any other such state to arbitrary precision.
A popular universal gate set is composed of the Hadamard ($H$), controlled-NOT (CNOT), and $T$ gates.\cite{Boykin.1999.10.1109/SFFCS.1999.814621}
The single-qubit $H$ gate creates superpositions and can be combined with the two-qubit CNOT gate to generate entangled states, \latin{i.e.}, states that cannot be expressed as a tensor product of single-qubit states.
The single-qubit $T$ gate is the fourth root of the Pauli $Z$ gate and, together with the $H$ gate, can approximate any desired single-qubit gate to arbitrary precision.

Universal gate sets, such as the one mentioned above, are extremely compact, as each element is at most a two-qubit gate.
However, the unconstrained use of gates from such universal sets may give rise to overly expressive circuits that explore the entire $n$-qubit Hilbert space, which is problematic in the context of quantum simulations of many-body systems.
Indeed, the starting point of such simulations is an easily prepared reference state that typically respects all symmetries of the target state, and ideally the employed quantum algorithm should preserve these symmetries as well.
Recall that symmetries can be represented by operators that commute with the Hamiltonian.
A suitable subset of these operators, together with the Hamiltonian, can be chosen as a complete set of commuting observables whose eigenvalues are conserved quantities and label the different symmetry sectors.
In the case of chemistry applications, for example, the electronic Hamiltonian conserves the particle number ($N$), the irreducible representation of the molecular point group, and, in a non-relativistic setting, the $z$-projection of the total spin ($S_z$) and the total spin squared ($S^2$).
Considering the above, the na\"{\i}ve use of a universal gate set whose elements generally violate symmetries may populate symmetry sectors outside the target one.
This issue is especially pronounced in variational quantum eigensolvers (VQEs),\cite{Peruzzo.2014.10.1038/ncomms5213} in which the optimization may collapse to a lower-energy state with undesired symmetry.

\begin{figure*}[htpb]
    \centering
    \includegraphics[width=6.5in]{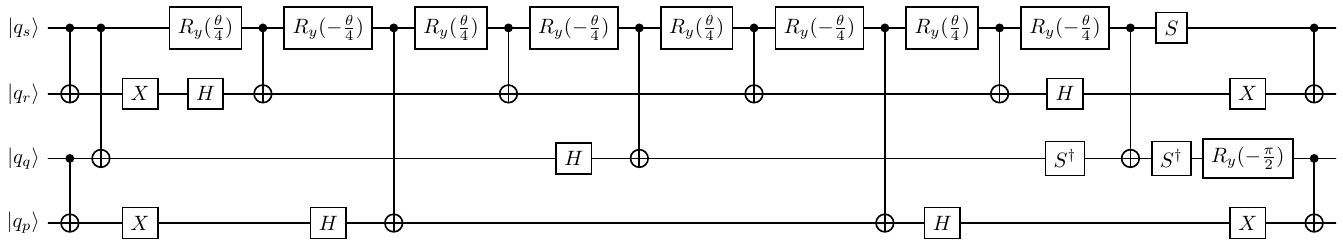}
    \caption{Quantum circuit performing a Givens rotation by an angle $\theta$ in the $\ket{1_p 1_q 0_r 0_s}$--$\ket{0_p 0_q 1_r 1_s}$ subspace of the four qubits labeled $p$, $q$, $r$, and $s$.
    }
    \label{fig:qeb}
\end{figure*}

To avoid exploring irrelevant symmetry sectors, one often replaces low-level universal gate sets with universal symmetry-adapted operator pools, which generate quantum circuits that can approximate any unitary transformation within the target symmetry sector of interest.
This is accomplished by requiring each element $P_i$ of a universal pool $\mathcal{P}$ to commute with the desired set of symmetry operators $\mathcal{S}$, \latin{i.e.}, $\mathcal{P} = \{P_i \mid \comm{P_i}{S_j}=0,\, \forall S_j \in \mathcal{S}\}$.
In typical applications to electronic structure and many-body physics, the pool elements are chosen to be symmetry-adapted many-body operators.
Each operator $P_i$ generates a unitary
\begin{equation}
    U_i (\theta_i) = e^{\theta_i P_i},
\end{equation}
with $\theta_i \in \mathbb{R}$.
In general, the generators $P_i$ can be either anti-Hermitian ($P_i = A_i$ with $A_i^\dagger = -A_i$), as is typically done in ansatz construction, or Hermitian ($P_i = -\mathrm{i} H_i$ with $H_i^\dagger = H_i$), reminiscent of time evolution and quantum phase estimation.\cite{Kitaev.1995.quant-ph/9511026}
At the hardware level, each unitary generated by an element of the operator pool $\mathcal{P}$ is implemented as a circuit of elementary single- and two-qubit gates with specific layouts that enforce the desired symmetries.
For example, a Givens rotation by an angle $\theta$ in the $\ket{1_p 1_q 0_r 0_s}$--$\ket{0_p 0_q 1_r 1_s}$ subspace of the qubits labeled $p$, $q$, $r$ and $s$ can be efficiently implemented by the quantum circuit shown in \cref{fig:qeb},\cite{Yordanov.2020.10.1103/PhysRevA.102.062612} which contains $H$, $X$, $S$, $S^\dagger$, $R_y$, and CNOT gates in an arrangement that conserves particle number and, with a careful choice of excitation indices, spatial symmetry and $S_z$ for any value of $\theta$.
However, enforcing symmetries at the gate level leads to longer and deeper quantum circuits.
Indeed, we have recently shown that a minimal universal operator pool that respects all symmetries relevant to chemistry contains elements whose circuit implementation requires more than 800 $R_y$ and 700 CNOT gates.\cite{Magoulas.2025.2511.13485}

Between universal gate sets that abandon all symmetries and fully symmetry-preserving operator pools lie various pools that enforce only a subset of symmetries.
Notable examples include (i) the qubit pool,\cite{Ryabinkin.2018.10.1021/acs.jctc.8b00932,Ryabinkin.2020.10.1021/acs.jctc.9b01084,Tang.2021.10.1103/PRXQuantum.2.020310} which respects $\mathbb{Z}_2$ symmetries; (ii) the simplified qubit-excitation-based\cite{Sun.2025.10.1021/acs.jctc.5c00119}, coupled-exchange-operator,\cite{Ramoa.2025.10.1038/s41534-025-01039-4} and generalized singles and doubles (GSD)\cite{Nooijen.2000.10.1103/PhysRevLett.84.2108,Nakatsuji.2000.10.1063/1.1287275} operator pools, which conserve spatial symmetry, $N$, and $S_z$; and (iii) singlet spin-adapted singles and perfect-pairing doubles (saGSpD),\cite{Anselmetti.2021.10.1088/1367-2630/ac2cb3,Burton.2023.10.1038/s41534-023-00744-2,Burton.2024.10.1103/PhysRevResearch.6.023300} which enforce $N$, $S_z$, and $S^2$.
Although enforcing spatial symmetry within saGSpD is conceptually straightforward, it has not been implemented in existing quantum algorithms.
Indeed, in the quantum-number-preserving (QNP) gate fabric\cite{Anselmetti.2021.10.1088/1367-2630/ac2cb3} and in tiled unitary product states (tUPS),\cite{Burton.2024.10.1103/PhysRevResearch.6.023300} the saGSpD gates are applied to adjacent qubits, necessarily violating point-group symmetry.
Meanwhile, the effect of enforcing spatial symmetry in saGSpD has not been investigated in the discretely optimized VQE (DISCO-VQE).\cite{Burton.2023.10.1038/s41534-023-00744-2}
It has been demonstrated numerically that iteratively constructed ans\"{a}tze based on these pools can reach chemical accuracy with very compact quantum circuits.
This is particularly true in the case of DISCO-VQE and tUPS, which combine the saGSpD pool with global optimization algorithms.
In this context, the interplay between universality, symmetry preservation, chemical accuracy, and the burden on quantum resources sheds new light on L\"{o}wdin's symmetry dilemma\cite{Lykos.1963.10.1103/RevModPhys.35.496} and brings it into the realm of quantum computing.

In this work, we highlight an underappreciated aspect of benchmarking quantum algorithms, \latin{i.e.}, that symmetry is generally neither explored in its full depth nor are the consequences of its absence systematically examined.
This can inadvertently lead to the design of very compact quantum algorithms that may suffer from variational collapse.
Here, we examine the dangers associated with symmetry breaking in quantum simulations of chemical systems.

On the theoretical front, we prove that enforcing spatial symmetry within the saGSpD operator pool renders it non-universal, revealing a trade-off between strict symmetry adaptation and ansatz expressiveness.
On the numerical side, we perform adaptive derivative-assembled pseudo-Trotter ansatz VQE\cite{Grimsley.2019.10.1038/s41467-019-10988-2,Tang.2021.10.1103/PRXQuantum.2.020310,Yordanov.2021.10.1038/s42005-021-00730-0} (ADAPT-VQE) simulations using operator pools with and without full symmetry enforcement.
Our goal is to determine when a given operator pool can be safely employed without suffering from variational collapse.
To that end, we study three physically relevant scenarios.
In the first, the target state is the global energy minimum for all examined geometries.
In this case, we find that simulations that initially violate symmetries eventually restore them, collapsing to the desired state.
In the second scenario, the two lowest-energy states cross and differ in multiple symmetries, \latin{e.g.}, spatial and spin symmetries.
Here, as long as the selected operator pool respects at least one of the symmetries not shared by the two states, the simulations can distinguish between them and avoid variational collapse.
Finally, we consider the case where the two lowest-energy states cross and differ only in a single symmetry, \latin{e.g.}, spatial or spin symmetry.
In this most challenging case, the simulations converge to the target state only when the employed operator pool enforces the symmetry property that distinguishes the two states.

We would like to emphasize that, while our numerical examples employ specific operator pools in the context of ADAPT-VQE, the phenomena we analyze are generic.
Indeed, any quantum algorithm that relies on symmetry-violating gates may suffer from symmetry contamination and, potentially, variational collapse.
Taken together, our comprehensive analysis provides a practical guide for the design and benchmarking of symmetry-preserving quantum circuits and deepens our understanding of symmetry adaptation in the age of quantum computing.

The paper is organized as follows.
In \cref{sec:pools}, we define the various operator pools considered in this study.
The lack of universality of the spatial-symmetry-preserving saGSpD pool is demonstrated in \cref{sec:saGSpD}.
The computational details of the ADAPT-VQE numerical simulations are provided in \cref{sec:comp}, the results are presented in \cref{sec:results}, and conclusions are given in \cref{sec:conc}.

\section{Operator Pool Definitions and Symmetry Properties}\label{sec:pools}

In this work, we define our operator pools in terms of elementary, anti-Hermitian many-body operators, \(\{A_k\}\).
Each such element generates a parametrized unitary \(\exp(\theta_k A_k)\) that can be used to construct the ansatz:
\begin{equation}\label{eq:ansatz}
    \ket*{\Psi} = \prod_k e^{\theta_k A_k} \ket*{\Phi},
\end{equation}
with the reference state $\ket*{\Phi}$ typically being a product state.
One of the most prominent operator pools, tracing its origins to classical electronic structure methods,\cite{Coester.1958.10.1016/0029-5582(58)90280-3,Coester.1960.10.1016/0029-5582(60)90140-1,Cizek.1966.10.1063/1.1727484,Cizek.1969.10.1002/9780470143599.ch2,Kutzelnigg.1977.10.1007/978-1-4757-0887-5_5,Kutzelnigg.1982.10.1063/1.444231,Bartlett.1989.10.1016/S0009-2614(89)87372-5,Szalay.1995.10.1063/1.469641,Taube.2006.10.1002/qua.21198,Cooper.2010.10.1063/1.3520564,Evangelista.2011.10.1063/1.3598471,Harsha.2018.10.1063/1.5011033,Filip.2020.10.1063/5.0026141,Freericks.2022.10.3390/sym14030494,Anand.2022.10.1039/d1cs00932j} is the GSD pool,\cite{Nooijen.2000.10.1103/PhysRevLett.84.2108,Nakatsuji.2000.10.1063/1.1287275} which contains anti-Hermitian fermionic single,
\begin{equation}\label{eq:gs}
    A_p^q = a^q a_p - a^p a_q,
\end{equation}
and double,
\begin{equation}\label{eq:gd}
    A_{pq}^{rs} = a^r a^s a_q a_p - a^p a^q a_s a_r,
\end{equation}
excitation operators.
In this notation, the $p$, $q$, $r$, and $s$ indices denote generic spinorbitals and $a^p$ ($a_p$) is the creation (annihilation) operator acting on the $p$\textsuperscript{th} spinorbital.
If each element of the GSD pool is allowed to appear multiple times in the ansatz [\cref{eq:ansatz}], each with an independent optimization parameter, the GSD pool is universal.\cite{Evangelista.2019.10.1063/1.5133059}
Furthermore, its efficient quantum-circuit implementation is enabled by the fermionic excitation-based framework.\cite{Yordanov.2020.10.1103/PhysRevA.102.062612}

The incorporation of various important molecular symmetries is straightforward in GSD.
Indeed, \cref{eq:gs,eq:gd} automatically conserve the number of particles by having the same number of creation and annihilation operators in a given operator string.
Additionally, spatial symmetry can be enforced by considering only excitations that belong to the totally symmetric representation of the given point group.
Furthermore, to enforce good $S_z$ quantum numbers, the excitation indices are constrained such that the same number of $s_z = -\frac{1}{2}$ ($\downarrow$) and $s_z = \frac{1}{2}$ ($\uparrow$) electrons are created as are annihilated.
To that end, single excitations are divided into two cases, namely,
\begin{equation}\label{eq:gs_u}
    A_{\Pu}^{\Qu} = a^{\Qu} a_{\Pu} - a^{\Pu} a_{\Qu}
\end{equation}
and
\begin{equation}\label{eq:gs_d}
    A_{\Pd}^{\Qd} = a^{\Qd} a_{\Pd} - a^{\Pd} a_{\Qd},
\end{equation}
while six cases arise for double excitations, namely,
\begin{equation}\label{eq:gd_uuuu}
    A_{\Pu\Qu}^{\Ru\Su} = a^{\Ru} a^{\Su} a_{\Qu} a_{\Pu} - a^{\Pu} a^{\Qu} a_{\Su} a_{\Ru},
\end{equation}
\begin{equation}\label{eq:gd_dddd}
    A_{\Pd\Qd}^{\Rd\Sd} = a^{\Rd} a^{\Sd} a_{\Qd} a_{\Pd} - a^{\Pd} a^{\Qd} a_{\Sd} a_{\Rd},
\end{equation}
\begin{equation}\label{eq:gd_udud}
    A_{\Pu\Qd}^{\Ru\Sd} = a^{\Ru} a^{\Sd} a_{\Qd} a_{\Pu} - a^{\Pu} a^{\Qd} a_{\Sd} a_{\Ru},
\end{equation}
\begin{equation}\label{eq:gd_uddu}
    A_{\Pu\Qd}^{\Rd\Su} = a^{\Rd} a^{\Su} a_{\Qd} a_{\Pu} - a^{\Pu} a^{\Qd} a_{\Su} a_{\Rd},
\end{equation}
\begin{equation}\label{eq:gd_duud}
    A_{\Pd\Qu}^{\Ru\Sd} = a^{\Ru} a^{\Sd} a_{\Qu} a_{\Pd} - a^{\Pd} a^{\Qu} a_{\Sd} a_{\Ru},
\end{equation}
and
\begin{equation}\label{eq:gd_dudu}
    A_{\Pd\Qu}^{\Rd\Su} = a^{\Rd} a^{\Su} a_{\Qu} a_{\Pd} - a^{\Pd} a^{\Qu} a_{\Su} a_{\Rd}.
\end{equation}
Here, the uppercase indices $P$, $Q$, $R$, and $S$ denote generic spatial orbitals.
The only symmetry broken by the GSD operator pool is $S^2$.

By taking linear combinations of spinorbital operators, it is possible to construct spin-adapted operators.
Indeed, the two single excitation operators of $A_{\Pu}^{\Qu}$ and $A_{\Pd}^{\Qd}$, in \cref{eq:gs_u,eq:gs_d} respectively, can be combined to give two spin-adapted operators: one singlet and one triplet.
For example, when a triplet spin-adapted operator acts on a closed-shell Slater determinant, it generates the corresponding low-spin triplet configuration state function, unless the determinant is annihilated by the action of the operator.
In a similar manner, the six double excitation operators of \cref{eq:gd_uuuu,eq:gd_dddd,eq:gd_udud,eq:gd_uddu,eq:gd_duud,eq:gd_dudu} can be combined to generate two singlet, three triplet, and one quintet spin-adapted operators.

To enforce $S^2$ symmetry, the GSD pool needs to be singlet spin-adapted.
Within the orthogonally spin-adapted formalism and adopting the coupling scheme in which the spins of spinorbitals appearing in the upper and lower indices are independently coupled,\cite{Paldus.1977.10.1002/qua.560110511,Paldus.1977.10.1063/1.434526,Adams.1979.10.1103/PhysRevA.20.1,Chiles.1981.10.1063/1.441643,Piecuch.1989.10.1002/qua.560360402,Geertsen.1991.10.1016/S0065-3276(08)60364-0,Piecuch.1992.10.1007/BF01113244,Piecuch.1994.10.1063/1.467304} the singlet spin-adapted GSD (saGSD) pool contains elements of the form
\begin{equation}\label{eq:sags}
A_P^Q = \frac{1}{\sqrt{2}} \left( A_{\Pu}^{\Qu} + A_{\Pd}^{\Qd} \right),
\end{equation}
\begin{equation}\label{eq:sagd_0}
\begin{split}
\tensor*[^{[0]}]{A}{_{PQ}^{RS}} = \frac{1}{2\sqrt{(1+\delta_{PQ})(1+\delta_{RS})}} &\left( A_{\Pu \Qd}^{\Ru \Sd} - A_{\Pu \Qd}^{\Rd \Su}\right.\\
&\left.- A_{\Pd \Qu}^{\Ru \Sd} + A_{\Pd \Qu}^{\Rd \Su}\right),
\end{split}
\end{equation}
and
\begin{equation}\label{eq:sagd_1}
	\begin{split}
	\tensor*[^{[1]}]{A}{_{PQ}^{RS}} ={}& \dfrac{(1-\delta_{PQ})(1-\delta_{RS})}{\sqrt{3}} \left[ A_{\Pu \Qu}^{\Ru \Su} + A_{\Pd \Qd}^{\Rd \Sd}\right.\\
	&\left.+ \frac{1}{2}\left( A_{\Pu \Qd}^{\Ru \Sd} + A_{\Pu \Qd}^{\Rd \Su} + A_{\Pd \Qu}^{\Ru \Sd} + A_{\Pd \Qu}^{\Rd \Su}\right) \right],
	\end{split}
\end{equation}
with the $\delta$ symbols denoting Kronecker deltas and the ``[0]'' and ``[1]'' superscripts designating intermediate spin quantum numbers.
These operators automatically enforce particle-number, $S_z$, and $S^2$ symmetry, while point-group symmetry can be enforced by only considering totally symmetric excitations.

Although the saGSD pool is universal and respects all symmetries relevant to chemistry, its quantum circuit implementation is exceedingly demanding in terms of quantum resources.
For example, we have recently shown that the efficient circuit representation of a unitary generated by a single operator of the form of \cref{eq:sagd_1} requires 8 CNOT staircases plus $\sim 3600$ CNOTs, and 3696 $R_y$ gates.\cite{Magoulas.2025.2511.13485}
In practice, it is therefore natural to consider truncated versions obtained by downselecting specific classes of spin-adapted operators.

One approximation to saGSD, recently introduced by our group, is the pDint0 operator pool.\cite{Magoulas.2025.2511.13485}
In this construction, the pool is restricted to perfect-pairing double excitations $A_{PP}^{QQ}$ [\cref{eq:sagd_0} with $P=Q$ and $R=S$] together with those double excitations that proceed through an intermediate singlet with no repeated upper and lower indices [\cref{eq:sagd_0} with $P\neq Q$ and $R\neq S$].
As a subset of saGSD, the pDint0 pool continues to conserve particle number, $S_z$, and $S^2$, and it can be chosen to respect spatial symmetry by including only totally symmetric excitations.
By removing operators of the form of \cref{eq:sagd_1}, the pDint0 pool is more practical than full saGSD.
Nevertheless, the computational cost remains substantial, with the circuit implementation of a single unitary requiring as much as 2 CNOT staircases plus $\sim 750$ CNOTs, and 864 $R_y$ gates.

Arguably, the approximation that leads to the most compact circuit representations is the saGSpD pool, which retains all spin-adapted single excitations [\cref{eq:sags}] and restricts the double excitations to the perfect-pairing ones, $A_{PP}^{QQ}$. 
Similar to pDint0, the saGSpD pool preserves particle number, $S_z$, and $S^2$, and spatial symmetry can again be enforced by selecting only totally symmetric excitations.
At the same time, the drastic reduction in the number and structure of double excitations leads to significantly more compact ans\"{a}tze and shallower circuits, making saGSpD particularly attractive for applications using current noisy quantum hardware.

\section{Non-Universality of the Totally Symmetric \NoCaseChange{saGSpD} Pool}\label{sec:saGSpD}

The saGSpD pool has a known limitation.\cite{Burton.2023.10.1038/s41534-023-00744-2}
If the chemical system under consideration is fully spin-polarized, \latin{i.e.}, all electrons have the same $s_z$ value, then unitaries generated by perfect-pairing operators act as the identity on the reference Slater determinant.
In this case, the only viable excitations are singles, rendering the optimization equivalent to Hartree--Fock (HF) theory due to the Thouless theorem.\cite{Thouless.1960.10.1016/0029-5582(60)90048-1}
The same issue can occur if all spinorbitals with a given $s_z$ value are fully occupied, although this scenario is highly unlikely in realistic applications that rely on larger-than-minimum basis sets.

Here we highlight another situation in which the universality of the saGSpD pool is compromised.
Specifically, for molecules with non-trivial point-group symmetry, \latin{i.e.}, belonging to point groups other than $C_1$, the saGSpD pool is not universal when spatial symmetry is enforced.
Conversely, when non-totally symmetric operators are incorporated into it, the saGSpD pool is universal but quantum simulations may suffer from symmetry breaking.
To demonstrate that the saGSpD pool is not universal when spatial symmetry is enforced, we will show that nested commutators of singlet spin-adapted singles and perfect-pairing doubles cannot generate all types of elements in the totally symmetric saGSD pool, which is universal.

To make this argument precise, we first recall that the standard Abelian point groups commonly used in classical electronic structure calculations, namely, $C_\text{i}$, $C_2$, $C_\text{s}$, $D_2$, $C_\text{2v}$, $C_\text{2h}$, and $D_\text{2h}$, are examples of elementary Abelian 2-groups, meaning they are Abelian and all elements other than the identity have order 2.
Thus, since perfect-pairing doubles $A_{PP}^{QQ}$ are totally symmetric by nature, enforcing spatial symmetry within the saGSpD pool amounts to restricting the spatial orbitals in spin-adapted singles $A_P^Q$ so that they belong to the same irreducible representation of the point group, $\Gamma_P = \Gamma_Q$.
In addition to these elements, the full saGSD pool contains double excitations of the form $A_{PP}^{QR}$ [\cref{eq:sagd_0} with $P=Q$], $\tensor*[^{[0]}]{A}{_{PQ}^{RS}}$, and $\tensor*[^{[1]}]{A}{_{PQ}^{RS}}$.
For excitations of the type $A_{PP}^{QR}$ to be totally symmetric, spatial orbitals $Q$ and $R$ must belong to the same irreducible representation of the point group, $\Gamma_Q = \Gamma_R$.
When it comes to the more complicated spin-adapted doubles, $\tensor*[^{[0]}]{A}{_{PQ}^{RS}}$ and $\tensor*[^{[1]}]{A}{_{PQ}^{RS}}$, there are five cases that result in totally symmetric operators, summarized in \cref{tab:tot_sym}.
For the totally symmetric saGSpD pool to be universal, nested commutators of its elements must generate all of these types of double excitation operators.

\begin{table*}[htbp]
  \centering
  \caption{\label{tab:tot_sym}
  Conditions for saGSD excitation operators to be totally symmetric in elementary Abelian 2-groups.}
  \begin{tabular}{ccc}
    \toprule
    Operator & Orbital index restrictions\textsuperscript{a} & Conditions for totally symmetric \\
    \midrule
    $A_P^Q$ & $P<Q$ & $\Gamma_P = \Gamma_Q$ \\[6pt]
    $A_{PP}^{QQ}$ & $P<Q$ & always totally symmetric \\[6pt]
    $A_{PP}^{QR}$ & $Q<R$ & $\Gamma_Q = \Gamma_R$ \\[6pt]
    \multirow{5}{*}{$\tensor*[^{[0,1]}]{A}{_{PQ}^{RS}}$} & \multirow{5}{*}{$P<Q, P\le R, \xi<S$\textsuperscript{b}} &
    $\Gamma_P = \Gamma_Q = \Gamma_R = \Gamma_S$ \\
    && $\Gamma_P = \Gamma_Q \neq \Gamma_R = \Gamma_S$ \\
    && $\Gamma_P = \Gamma_R \neq \Gamma_Q = \Gamma_S$ \\
    && $\Gamma_P = \Gamma_S \neq \Gamma_Q = \Gamma_R$ \\
    && $\Gamma_P, \Gamma_Q, \Gamma_R, \Gamma_S$ distinct and
    $\Gamma_P \otimes \Gamma_Q \otimes \Gamma_R \otimes \Gamma_S = \Gamma_\text{tot}$\textsuperscript{c}\\
    \bottomrule
  \end{tabular}
  
  \textsuperscript{a}The index restrictions ensure that only unique excitations are considered.
  \textsuperscript{b}$\xi = Q$ if $R = P$, otherwise $\xi = R$.
  \textsuperscript{c}$\Gamma_\text{tot}$ denotes the totally symmetric irreducible representation of the point group.
\end{table*}

We begin by noting that spin-adapted singles and perfect-pairing doubles preserve the parity of the number of electrons in a given irreducible representation $\mathcal{I}$ ($\Pi_\mathcal{I}$), defined as
\begin{equation}
    \Pi_\mathcal{I} = \prod_{\left\{P \mid \Gamma_P = \mathcal{I}\right\}} (1-2n_{\Pu})(1-2n_{\Pd}).
\end{equation}
Indeed, spin-adapted singles promote one electron between spatial orbitals of the same irreducible representation, leaving the number of electrons in that irreducible representation unchanged.
Similarly, perfect-pairing doubles at most reduce the number of electrons in one irreducible representation by two and increase the number of electrons in another by two, so that the number parities in these irreducible representations remain invariant.
Considering that commutators of saGSpD operators give rise to linear combinations of such parity-preserving operators, all elements of the Lie algebra generated by the totally symmetric saGSpD pool preserve the number parity of every irreducible representation.
Therefore, double excitations of the form $\tensor*[^{[0]}]{A}{_{PQ}^{RS}}$ and $\tensor*[^{[1]}]{A}{_{PQ}^{RS}}$ with $\Gamma_P$, $\Gamma_Q$, $\Gamma_R$, and $\Gamma_S$ distinct, which necessarily break the number parity in some irreducible representations, do not belong to the Lie algebra generated by the totally symmetric saGSpD pool.
This demonstrates the lack of universality of the totally symmetric saGSpD pool for Abelian point groups with at least four distinct irreducible representations, such as $D_2$, $C_\text{2v}$, $C_\text{2h}$, and $D_\text{2h}$.

To illustrate the lack of universality for Abelian point groups with fewer than four irreducible representations, such as $C_\text{i}$, $C_2$, and $C_\text{s}$, we examine under which conditions nested commutators involving spin-adapted singles and perfect-pairing doubles can generate excitations of the form $A_{\Pu\Qu}^{\Ru\Su}$ and $A_{\Pd\Qd}^{\Rd\Sd}$, which appear in the definition of $\tensor*[^{[1]}]{A}{_{PQ}^{RS}}$.
We begin by noting that, in their pure form, such excitation operators do not belong to the Lie algebra generated by the saGSpD operators.
This follows from two observations.
First, although nonzero commutators involving the $A_P^Q$ operators preserve the many-body rank, they give rise to linear combinations of operators with one spinorbital index replaced by another of the same $s_z$ value.
This already implies that nested commutators that generate the excitations $A_{\Pu\Qu}^{\Ru\Su}$ and $A_{\Pd\Qd}^{\Rd\Sd}$ must involve at least two perfect-pairing doubles.
Second, nonzero commutators involving even a two-body operator and perfect-pairing doubles result in operators whose many-body rank is increased by 1.
Therefore, $A_{\Pu\Qu}^{\Ru\Su}$ and $A_{\Pd\Qd}^{\Rd\Sd}$ can only appear as a product with strings of number operators.

Indeed, as shown in the \sm, one possible route to generate these operators is
\begin{equation}
\begin{split}
    \comm{\comm{\comm{A_{PP}^{QQ}}{A_Q^R}}{A_{QQ}^{SS}}}{A_P^S} ={}\\
    \frac{1}{2}\left( A_{\Pd\Qd}^{\Rd\Sd} + A_{\Pd\Qu}^{\Ru\Sd} \right) \left( n_{\Pu} - n_{\Su} \right)\\
    + \frac{1}{2}\left( A_{\Pu\Qu}^{\Ru\Su} + A_{\Pu\Qd}^{\Rd\Su} \right) \left( n_{\Pd} - n_{\Sd} \right).
\end{split}
\end{equation}
Note that the number-operator factors ensure that both terms will either be zero or applied simultaneously, rendering the entire expression spin-adapted.
If we assume that spatial orbital $P$ is doubly occupied and $S$ unoccupied, the above commutator will be proportional to
\begin{equation}\label{eq:linear_comb}
    \begin{split}
        \comm{\comm{\comm{A_{PP}^{QQ}}{A_Q^R}}{A_{QQ}^{SS}}}{A_P^S} \propto\\
        A_{\Pd\Qd}^{\Rd\Sd} + A_{\Pd\Qu}^{\Ru\Sd} + A_{\Pu\Qu}^{\Ru\Su} + A_{\Pu\Qd}^{\Rd\Su}\\
        = \sqrt{3} \tensor*[^{[1]}]{A}{_{PQ}^{RS}} - \tensor*[^{[0]}]{A}{_{PQ}^{RS}}.
    \end{split}
\end{equation}
Recall that single excitations impose restrictions on the irreducible representations of the orbitals to ensure that the overall excitation is totally symmetric.
These restrictions appear between orbitals involving the upper and lower indices of the final result.
In this particular example, the restrictions are $\Gamma_P = \Gamma_S$ and $\Gamma_Q = \Gamma_R$.
As a result, the Lie algebra of the totally symmetric saGSpD pool cannot generate excitations $\tensor*[^{[1]}]{A}{_{PQ}^{RS}}$ with $\Gamma_P = \Gamma_Q$, $\Gamma_R = \Gamma_S$, and $\Gamma_P \neq \Gamma_R$.
This demonstrates the lack of universality of saGSpD in situations involving only two irreducible representations, including the $C_\text{i}$, $C_2$, and $C_\text{s}$ point groups.
The above analysis also implies that the more irreducible representations appear in the orbital picture of a molecular system, the more severe the deficiency of the saGSpD pool will be, as more types of excitations will become unattainable.

In concluding this section, we would like to point out another possible factor contributing to the lack of universality of the totally symmetric saGSpD pool.
As shown in \cref{eq:linear_comb} and the \sm, nested commutators of spin-adapted singles and perfect-pairing doubles typically generate linear combinations of spin-adapted operators.
Consequently, when one employs the Baker--Campbell--Hausdorff identity to express a product of unitaries generated by saGSpD operators as a single exponential, many of the resulting operators will be linearly dependent.
It is then possible that the spatial-symmetry restrictions may prevent overcoming some of these linear dependencies, resulting in a number of linearly independent operators that is smaller than the dimension of the symmetry-adapted full configuration interaction (FCI) problem.

\section{Computational Details}\label{sec:comp}

To illustrate the conditions under which symmetry-breaking pools can be safely employed in quantum algorithms, we numerically examined three physically relevant scenarios.
In the first scenario, a single electronic state is the global energy minimum along a dissociation profile.
This situation is modeled by the $D_{\infty\text{h}}$-symmetric dissociation of the \ce{H6} linear chain, a prototypical strongly correlated model system.
In the second scenario, as the nuclear geometry is continuously modified, the lowest-energy states of two symmetry sectors cross and differ in multiple symmetry properties.
An illustrative example of this process is the $C_\text{2v}$-symmetric bending mode of methylene, \ce{CH2}, where the lowest-energy $\tensor*[^{3}]{B}{_{1}}$ and $\tensor*[^{1}]{A}{_{1}}$ states cross.
When we focus on the low-spin component of $\tensor*[^{3}]{B}{_{1}}$, these two states belong to different irreducible representations of the point group and have different spin multiplicities.
In the third scenario, the lowest-energy states of two symmetry sectors cross but differ in only a single symmetry property.
We consider two molecular processes that exhibit this behavior.
The first is the $C_\text{s}$-symmetric bending mode in an asymmetrically stretched \ce{BeH2} molecule, where the lowest-energy $\tensor*[^{1}]{A}{^{\prime}}$ and $\tensor*[^{3}]{A}{^{\prime}}$ states cross.
These states share all symmetry properties except $S^2$ when we focus on the low-spin component of $\tensor*[^{3}]{A}{^{\prime}}$.
The second is the $C_{\infty\text{v}}$-symmetric dissociation of the \ce{BO} diatomic, in which the lowest-energy $\tensor*[^{2}]{\Sigma}{^{+}}$ and $\tensor*[^{2}]{\Pi}{}$ states cross.
These states differ only in the irreducible representation of the point group.

In our numerical simulations, we considered three types of operator pools: (1) pools that are universal but break symmetries, (2) pools that enforce all symmetries but are not universal, and (3) pools that enforce all symmetries and are universal.
Specifically, for the first category we employed the GSD pool, which breaks $S^2$ symmetry, and a variant of the saGSpD pool that includes spatial-symmetry-violating singles.
The saGSpD pool with spatial symmetry enforced was used as a representative of the second type.
To distinguish between the two variants of saGSpD, we refer to the spatial-symmetry-violating variant as saGSpD-full throughout the article.
For the third type of pools, we used the pDint0 pool, which is, to the best of our knowledge, the most circuit-efficient, fully symmetry-adapted universal operator pool.

\begin{table}[htbp]
    \centering
    \caption{\label{tab:pools}
    Operator pools employed in this work, the symmetries they conserve, and their universality.}
    \begin{tabular}{c c c c c c}
    \toprule
    \multirow{2}{*}{Pool} & \multicolumn{4}{c}{\underline{Conserved Quantities}} & \multirow{2}{*}{Universal}\\
    & $N$ & $\Gamma$\textsuperscript{a} & $S_z$ & $S^2$ &  \\
    \midrule 
    GSD &  \checkmark & \checkmark & \checkmark &  & \checkmark \\
    saGSpD-full & \checkmark &  & \checkmark & \checkmark & \checkmark \\
    saGSpD & \checkmark & \checkmark & \checkmark & \checkmark &  \\
    pDint0 & \checkmark & \checkmark & \checkmark & \checkmark & \checkmark \\
    \bottomrule
    \end{tabular}
    
    \textsuperscript{a}$\Gamma$ denotes the irreducible representation of the point group.
\end{table}

The quantum algorithm we employed was ADAPT-VQE,\cite{Grimsley.2019.10.1038/s41467-019-10988-2} which uses the following two-step process to iteratively construct and optimize the wavefunction ansatz.
First, the ansatz is expanded by appending the new operator(s) from a predefined pool based on the selection criterion of choice.
Second, all parameters are optimized variationally in a conventional VQE step.
The ansatz-expansion and parameter-optimization cycles repeat until a convergence threshold is met.

In the ADAPT-VQE simulations with the GSD, saGSpD, and pDint0 pools, at each macro-iteration the operator with the largest energy-gradient magnitude was selected from the pool and added to the ansatz.
For ADAPT-VQE-saGSpD-full, the energy gradient of non-totally symmetric singles is zero by symmetry.
To overcome this issue, the pool was extended by adding elements formed from ordered tuples of two and three spin-adapted singles, inspired by the gate patterns observed in the QNP and tUPS approaches.
To be precise, the tuples were constructed from non-totally symmetric singles such that the direct product of the irreducible representations of the individual generators is totally symmetric.
Using the $C_\text{2v}$ point group as an example, one such tuple is $(A_P^Q, A_R^S)$ with both spin-adapted singles belonging to the $B_1$ irreducible representation.
Using the closed-form expression for unitaries generated by singlet spin-adapted singles,\cite{Magoulas.2025.10.1080/00268976.2025.2534672} it is straightforward to show that the corresponding product of unitaries contains both $A_1$ and $B_1$ components.
When such tuples were selected by the ADAPT-VQE algorithm, the ansatz was expanded by appending to it the product of the unitaries generated by each of the operators in the tuple, with each unitary associated with an independent optimization parameter.
The tuples were formed such that the product of the irreducible representations of the individual elements was totally symmetric.
Furthermore, we employed a more greedy selection criterion based on steepest energy descent via simplicial homology global optimization (SHGO)\cite{Endres.2018.10.1007/s10898-018-0645-y} as implemented in \textsc{SciPy}.\cite{Virtanen.2020.10.1038/s41592-019-0686-2} The number of sampling points used in the construction of the simplicial complex was set to 32 using the Sobol' sampling method.\cite{Joe.2008.10.1137/070709359} The amplitude of the new operator(s) was initialized to the SHGO result for the following VQE optimization.

The simulations were terminated once the number of ansatz parameters equaled the dimension of the pertinent symmetry-adapted FCI problem minus one for normalization or when the simulation stagnated, repeatedly selecting the same operator.
In the VQE steps, we used the Broyden--Fletcher--Goldfarb--Shanno\cite{Broyden.1970.10.1093/imamat/6.3.222,Fletcher.1970.10.1093/comjnl/13.3.317,Goldfarb.1970.10.1090/S0025-5718-1970-0258249-6,Shanno.1970.10.1090/S0025-5718-1970-0274029-X} (BFGS) optimizer, as implemented in \textsc{SciPy}, with a tight convergence criterion of \SI{e-6}{\hartree} for the norm of the gradient vector and a maximum of 2000 micro-iterations. 
However, na\"{i}ve BFGS optimization did not generate tightly converged energetics for the dissociation of BO with the saGSpD-full pool.
To that end, we applied the basin-hopping global optimization technique\cite{Wales.2005.10.1088/1478-3975/2/4/S02} as implemented in \textsc{SciPy}, setting the maximum number of iterations to 30 and the temperature, which controls the probability of accepting a new state, to \SI{e-4}{\hartree}.

All numerical simulations were based on restricted (closed- or open-shell) HF references.
The one- and two-electron integrals were obtained with \textsc{Psi4}.\cite{Smith.2020.10.1063/5.0006002}
The STO-6G basis set\cite{Hehre.1969.10.1063/1.1672392} was used throughout, and all orbitals were correlated in the ADAPT-VQE simulations, except in the case of BO where the four lowest-energy $\sigma$ orbitals, correlating with the 1s and 2s atomic orbitals of the B and O atoms, were frozen.
To ensure a smooth HF potential for BO, we proceeded as follows.
First, we optimized the HF reference for the $^2\Sigma^+$ state at $R_\text{B--O} = \SI{1.2}{\angs}$, and used the resulting orbitals in an HF calculation at $R_\text{B--O} = \SI{1.5}{\angs}$.
This process was repeated for the \SI{1.8}{\angs} and \SI{2.1}{\angs} internuclear distances.

The saGSpD, saGSpD-full, and pDint0 operator pools were implemented in a local version of the \textsc{QForte} package\cite{Stair.2022.10.1021/acs.jctc.1c01155}.

\section{Results}\label{sec:results}
\subsection{Scenario I: Global Ground State}
\label{subsec:H6}

We begin with the best-case scenario, in which the lowest-energy state of the target symmetry sector is the global energy minimum.
In a minimum-basis-set description, linear \ce{H6} has six spatial orbitals whose irreducible representations alternate between $A_\text{g}$ and $B_\text{1u}$ when the $D_\text{2h}$ point group is employed, the largest Abelian subgroup of the full $D_{\infty \text{h}}$ symmetry group.
As a result, the only possible molecular term symbols characterizing the potential energy curves (PECs) arising from the $D_{\infty \text{h}}$-symmetric dissociation of \ce{H6}/STO-6G are $\tensor*[]{\Sigma}{^{+}_{\text{g}}}$, with spin multiplicities ranging from singlet to quintet, and $\tensor*[]{\Sigma}{^{+}_{\text{u}}}$, with spin multiplicities ranging from singlet to septet.

\begin{figure*}[htpb]
    \centering
    \includegraphics[width=6in]{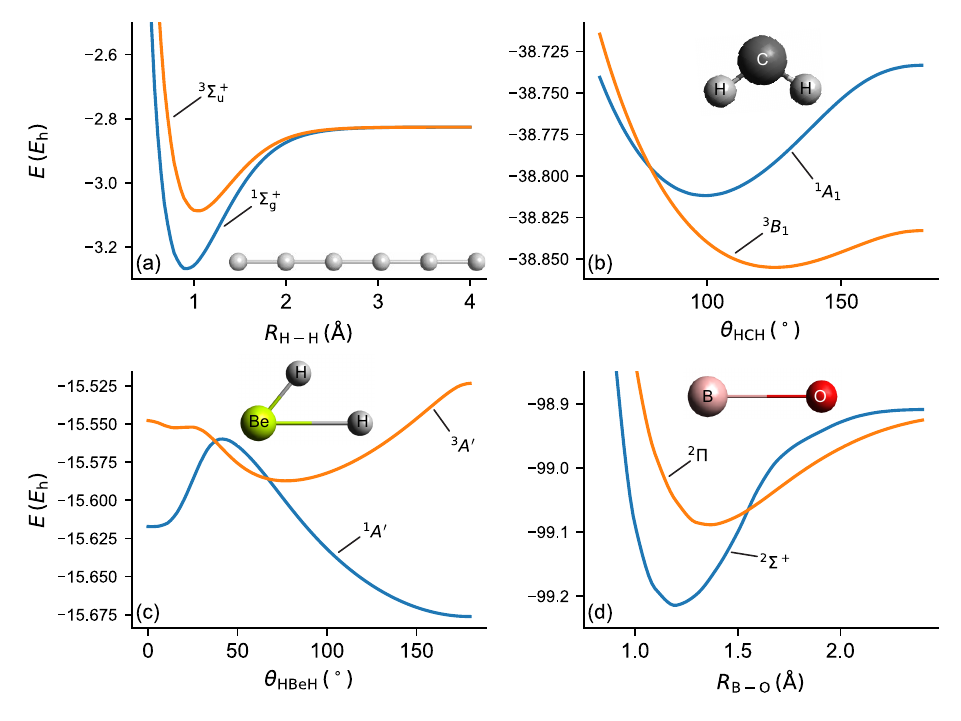}
    \caption{The two lowest-energy potential energy curves for the systems of interest, computed at the FCI/STO-6G level of theory: (a) $D_{\infty\text{h}}$-symmetric dissociation of \ce{H6}; (b) $C_\text{2v}$-symmetric bending of \ce{CH2}; (c) $C_\text{s}$-symmetric bending of asymmetrically stretched \ce{BeH2}; and (d) $C_{\infty \text{v}}$-symmetric dissociation of BO.
    }
    \label{fig:fci}
\end{figure*}

\begin{figure}[htpb]
    \centering
    \includegraphics[width=3.37in]{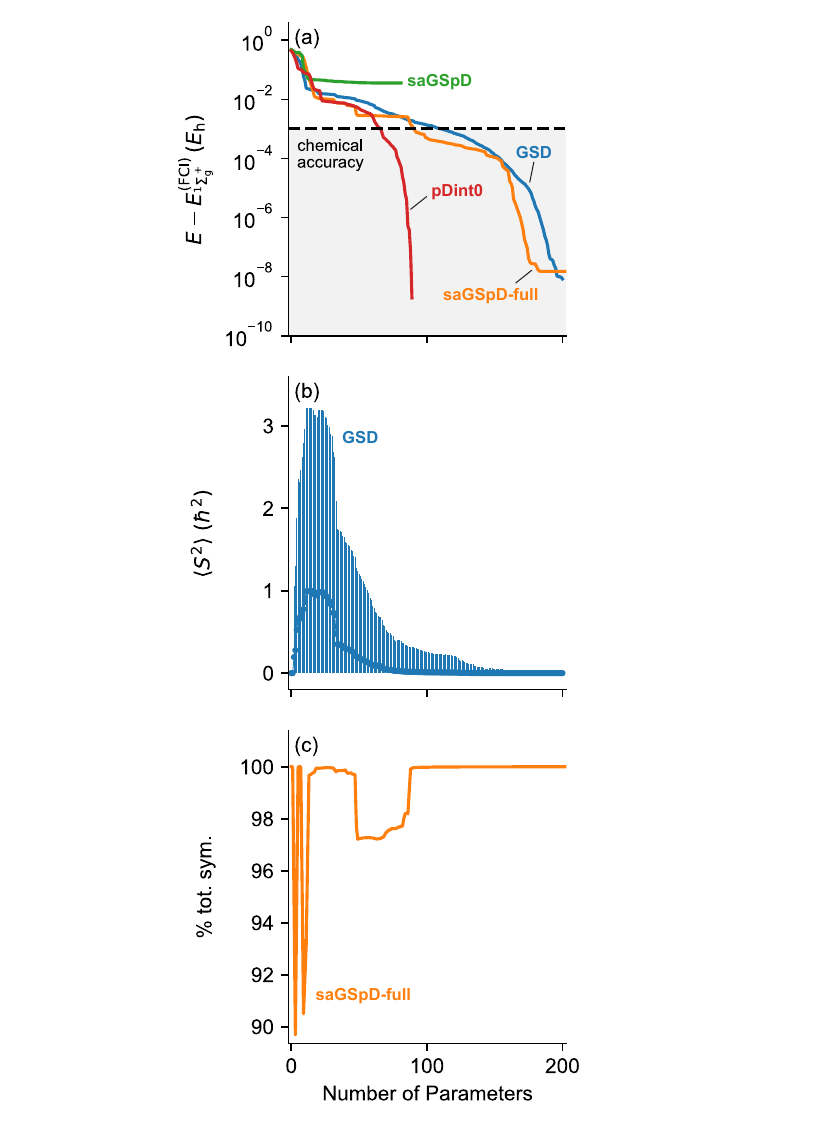}
    \caption{Convergence of ADAPT-VQE/STO-6G simulations using the GSD, saGSpD-full, saGSpD, and pDint0 operator pools for the \ce{H6} linear chain at the $R_{\text{H--H}}=\SI{2.0}{\angs}$ geometry, shown as functions of the number of parameters:
(a) Energy errors relative to the FCI/STO-6G energy of the lowest-energy $\tensor*[^{1}]{\Sigma}{^{+}_{\text{g}}}$ state;
(b) expectation values of the total spin squared operator and their standard deviations using the GSD pool; and
(c) weight of the totally symmetric component using the saGSpD-full pool. 
}
    \label{fig:h6}
\end{figure}

As shown in panel (a) of \cref{fig:fci}, the two lowest-energy states of \ce{H6}/STO-6G are of $\tensor*[^{1}]{\Sigma}{^{+}_{\text{g}}}$ and $\tensor*[^{3}]{\Sigma}{^{+}_{\text{u}}}$ symmetries.
Since $\tensor*[^{1}]{\Sigma}{^{+}_{\text{g}}}$ is the global energy minimum for all examined geometries, even algorithms based on symmetry-breaking operator pools may successfully target this state, as long as the pool is universal.
Indeed, as shown in \cref{fig:h6}, all ADAPT-VQE simulations paired with universal operator pools, with and without full symmetry enforced, produced numerically exact results for \ce{H6}/STO-6G with an internuclear distance between neighboring hydrogen atoms of $R_\text{H--H} = \SI{2.0}{\angs}$.
In what follows, we discuss in detail the performance of the various operator pools.

We begin with the pDint0 operator pool that is universal and enforces all symmetries.
Although its hardware implementation requires substantially longer and deeper circuits, it provides a benchmark to assess the performance of the more economical, symmetry-violating pools employed in this study.
As shown in panel (a) of \cref{fig:h6} for the $R_\text{H--H} = \SI{2.0}{\angs}$ geometry of linear \ce{H6}/STO-6G, ADAPT-VQE-pDint0 reaches chemical accuracy, defined as a \SI{1}{\millihartree} error with respect to FCI, once 67 operators are incorporated to the ansatz.
By adding 20 more operators, ADAPT-VQE-pDint0 produces numerically exact results, \latin{i.e.}, within \SI{1.0}{\microhartree} from FCI.
This is consistent with the dimension of the fully symmetry-adapted Hilbert space, which is spanned by 92 six-electron, totally symmetric, singlet configuration state functions (CSFs).

Next, we consider the standard GSD operator pool that is universal and enforces all symmetries except $S^2$.
As anticipated, the $S^2$ symmetry is severely broken within the first few ADAPT-VQE macro-iterations.
As shown in panel (b) of \cref{fig:h6}, the expectation value of $S^2$ reaches a maximum of $\SI{1.0}{\hbar^2}$ with a standard deviation on the order of \SI{2.0}{\hbar^2} once 11 operators have been added to the ansatz.
Upon computing the overlap of the ADAPT-VQE wavefunction with the exact eigenvectors, we find that the largest symmetry contaminants at this stage of the simulation are $\ket{\tensor*[^{5}]{\Psi}{_{3}}}$, $\ket{\tensor*[^{5}]{\Psi}{_{6}}}$, $\ket{\tensor*[^{5}]{\Psi}{_{9}}}$, and $\ket{\tensor*[^{3}]{\Psi}{_{1}}}$, listed in decreasing order of weight (see Fig.\ S2 in the \sm).
To be precise, the wavefunction at this point has 16\% quintet and 0.5\% triplet character, explaining the observed large standard deviations.
As the simulation progresses, the contribution of the symmetry contaminants is suppressed while that of the ground-state eigenvector gradually increases to 1.0.
Similar observations hold when we focus on the weakly correlated equilibrium region and the strongly correlated dissociation limit, as shown in Figs.\ S1 and S3 in the \sm, respectively.
However, it is worth mentioning that although ADAPT-VQE-GSD is capable of producing numerically exact results for \ce{H6}/STO-6G, the convergence is very slow compared to its pDint0 counterpart.
In fact, ADAPT-VQE-GSD requires more than 100 parameters to reach chemical accuracy, while the dimension of the fully symmetry-adapted Hilbert space is 92.
This is due to the fact that ADAPT-VQE-GSD explores the much larger Hilbert space of dimension 200, spanned by six-electron, totally symmetric, $S_z = \SI{0}{\hbar}$ Slater determinants.

Subsequently, we proceed to the results obtained with the saGSpD-full operator pool that is universal, albeit sacrificing point-group symmetry.
Similar to the GSD case, at the early stages of the ADAPT-VQE-saGSpD-full simulation, spatial symmetry is substantially broken, with the wavefunction attaining an approximately 10\% $B_\text{1u}$ character, as shown in \cref{fig:h6}(c).
As the simulation progresses, the $A_\text{g}$ character is restored and the $\tensor*[^{1}]{\Sigma}{^{+}_{\text{g}}}$ state is obtained.
When considering the entire simulation, it is worth noting that the symmetry breaking appears to be less severe than when the GSD operator pool was employed.
In light of our analysis in \cref{sec:saGSpD}, this observation can be attributed to the fact that only two irreducible representations appear in a minimum-basis-set description of linear \ce{H6}, despite the use of the $D_\text{2h}$ point-group symmetry.
Finally, the ADAPT-VQE-saGSpD-full simulation exhibits a similarly slow convergence to FCI, as the corresponding Hilbert space is spanned by 175 six-electron singlet CSFs.

We conclude this subsection by discussing the ADAPT-VQE results obtained with the fully symmetry-adapted saGSpD operator pool.
As the linear \ce{H6} system has a symmetry higher than $C_1$, the saGSpD pool is not universal (see \cref{sec:saGSpD}).
This has a detrimental effect on the performance of ADAPT-VQE, which, although preserving all symmetries, does not reach chemical accuracy, and the simulations stagnate for all of the examined geometries (see \cref{fig:h6} and Figs.\ S4 and S8 in the \sm).
It is worth mentioning that the \SI{1.3}{\millihartree} error for the strongly correlated $R_\text{H--H} = \SI{3.0}{\angs}$ configuration is substantially smaller than the 24 and \SI{36}{\millihartree} deviations at 1.0 and \SI{2.0}{\angs}, respectively.
This is consistent with the observation that the classical coupled-cluster with perfect-pairing doubles approach provides numerically exact results for dissociating linear chains of hydrogen atoms.\cite{Henderson.2014.10.1063/1.4904384}
This seemingly surprising result can be explained by examining the contributions to the converged ADAPT-VQE-saGSpD state of the exact eigenvectors.
As shown in the \sm, for $R_\text{H--H} = \SI{3.0}{\angs}$, the ADAPT-VQE-saGSpD state is essentially a linear combination of the first four singlet eigenvectors, all of which converge to the same dissociation limit.
At that limit, all four states are degenerate and any linear combination of them is an eigenstate of the Hamiltonian with the same eigenvalue.

To showcase the lack of universality of the saGSpD operator pool for this numerical example, we computed the weights of the singlet CSFs spanning the pertinent Hilbert space in the ADAPT-VQE-saGSpD wavefunction.
Across all examined geometries, 2 out of the 92 singlet CSFs always remain orthogonal to the ADAPT-VQE-saGSpD wavefunction (see Figs.\ S5, S7, and S9 in the \sm).
The first one is obtained from the RHF Slater determinant ($\ket{\Phi}$) via a double excitation, going through an intermediate triplet, from the first (0) and third (2) occupied spatial orbitals to the first (3) and third (5) unoccupied ones, $\ket{\Phi_{0\,2}^{3\,5}}_1 = \tensor*[^{[1]}]{A}{_{0\,2}^{3\,5}} \ket{\Phi}$, with the subscript ``1'' in the CSF denoting the common intermediate spin quantum number.
The second orthogonal CSF is $\ket{\Phi_{1\,1\,0\,2}^{4\,4\,3\,5}}_{0,\frac{1}{2},1}$, with the subscript denoting a sequence of intermediate spin quantum numbers.
This CSF can be obtained from the RHF Slater determinant via a sequence of two excitations acting on distinct indices, namely, $\ket{\Phi_{1\,1\,0\,2}^{4\,4\,3\,5}}_{0,\frac{1}{2},1} = \tensor*[^{[1]}]{A}{_{0\,2}^{3\,5}} A_{1\,1}^{4\,4} \ket{\Phi}$.
Because spatial orbitals with even (odd) indices belong to the $A_\text{g}$ ($B_\text{1u}$) irreducible representation, the $\tensor*[^{[1]}]{A}{_{0\,2}^{3\,5}}$ excitation has $\Gamma_0 = \Gamma_2$, $\Gamma_3 = \Gamma_5$, and $\Gamma_0 \neq \Gamma_3$.
As shown in \cref{sec:saGSpD}, this is exactly one of the types of excitation operators absent from the Lie algebra of the saGSpD operator pool.
It is interesting to note that, although only two of the 92 CSFs remain orthogonal to it, the ADAPT-VQE-saGSpD simulations stagnate much earlier, repeatedly appending the same operator to the ansatz for the remainder of the simulations (see Figs.\ S4, S6, and S8 in the \sm). 
This is potentially a manifestation of the dimensionality deficit discussed earlier.
In passing, we note that one approach to overcome the limitations of the saGSpD pool without breaking any symmetries would be to pair it with non-iterative energy corrections.
Unfortunately, our preliminary numerical results employing a spin-adapted variant of the moment energy correction\cite{Magoulas.2023.10.1021/acs.jpca.3c02781} for the $R_\text{H--H} = \SI{2.0}{\angs}$ geometry of the \ce{H6}/STO-6G linear chain showed that, although the deviation from FCI is substantially reduced, it still does not reach chemical accuracy.

\subsection{Scenario II: Crossing of States with Multiple Symmetry Differences}
\label{subsec:CH2}

Now we turn our attention to the more challenging situation in which the lowest-energy state of the target symmetry sector crosses another state with distinct irreducible representations for two or more symmetries (\latin{e.g.}, spin and point group).
The specific molecular example we consider is the $C_\text{2v}$-symmetric bending mode of \ce{CH2}.
The carbon atom is placed at the origin in the $yz$ plane, the C--H internuclear distances are fixed at $R_\text{C--H} = \SI{1.117}{\angs}$, and the hydrogen atoms are arranged such that they share the $z$ coordinate and their $y$ coordinates differ only in sign.
In panel (b) of \cref{fig:fci}, we show the two lowest-energy FCI/STO-6G PECs of \ce{CH2} for $\theta_\text{HCH}$ values between $60^\circ$ and $180^\circ$.
These states have $\tensor*[^{1}]{A}{_{1}}$ and $\tensor*[^{3}]{B}{_{1}}$ symmetries and cross around $\theta_\text{HCH} = 80^\circ$.
As a result, the ground electronic state switches from $\tensor*[^{3}]{B}{_{1}}$ for $80^\circ \le \theta_\text{HCH} \le 180^\circ$ to $\tensor*[^{1}]{A}{_{1}}$ when $60^\circ \le \theta_\text{HCH} \le 80^\circ$.
At the linear geometry ($\theta_\text{HCH} = 180^\circ$), the $\tensor*[^{3}]{B}{_{1}}$ state correlates with the $\tensor*[^{3}]{\Sigma}{_{g}^{-}}$ state ($B_\text{2g}$ in $D_\text{2h}$), while the $\tensor*[^{1}]{A}{_{1}}$ state correlates with the $\tensor*[^{1}]{\Delta}{_{g}}$ state ($A_\text{g}$ component in $D_\text{2h}$).
To target the $\tensor*[^{1}]{A}{_{1}}$ state, even when the $\tensor*[^{3}]{B}{_{1}}$ state lies lower in energy, one needs to enforce either point-group or $S^2$ symmetry (or both).
We performed ADAPT-VQE/STO-6G simulations for $\theta_\text{HCH} = 60^\circ$ and $180^\circ$.
These representative geometries were chosen because at these bond angles the singlet--triplet gap is largest and the identity of the lowest-energy state flips.

\begin{figure*}[t]
    \centering
    \includegraphics[width=5in]{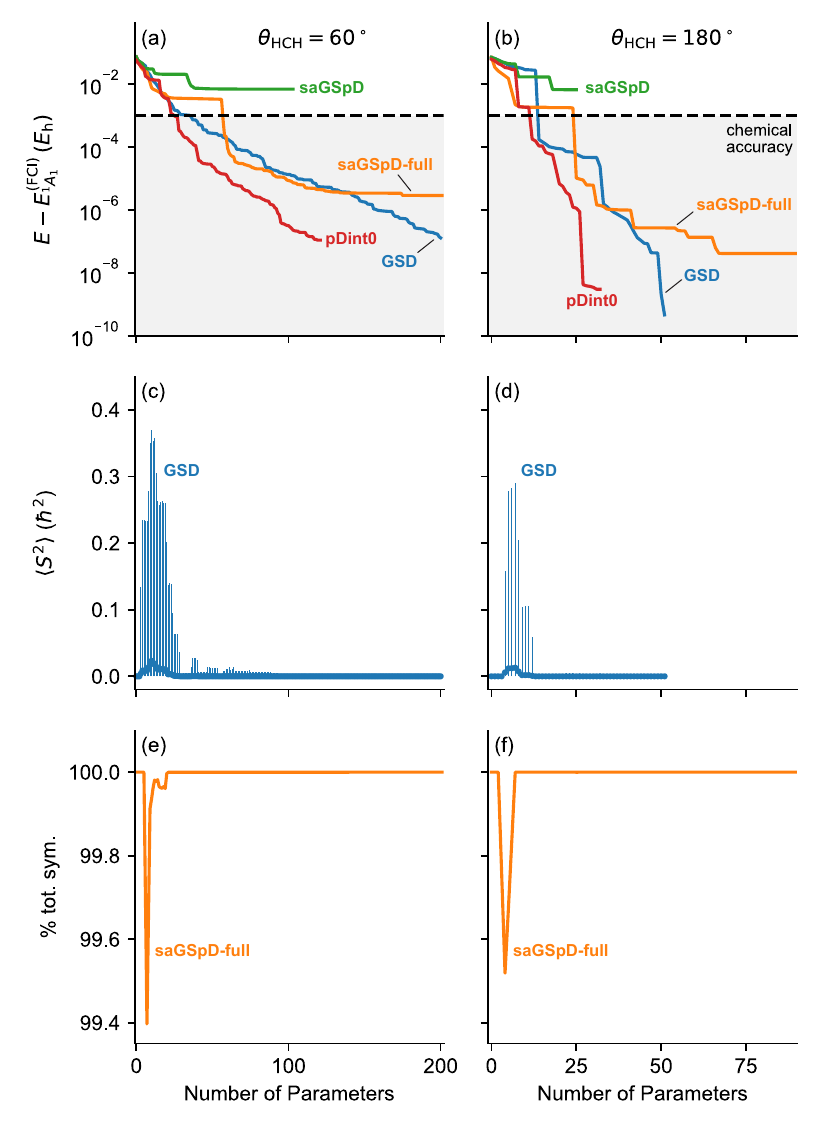}
    \caption{Convergence of ADAPT-VQE/STO-6G simulations using the GSD, saGSpD-full, saGSpD, and pDint0 operator pools for the \ce{CH2} molecule at the $\theta_{\text{HCH}}= 60^\circ$ and $180^\circ$ geometries, shown as functions of the number of parameters:
(a,b) Energy errors relative to the FCI/STO-6G energy of the lowest-energy $\tensor*[^{1}]{A}{_{1}}$ state ($\tensor*[^{1}]{A}{_{\text{g}}}$ for the linear geometry);
(c,d) expectation values of the total spin squared operator and their standard deviations using the GSD pool; and
(e,f) weight of the totally symmetric component using the saGSpD-full pool.
    }
    \label{fig:ch2}
\end{figure*}

As shown in \cref{fig:ch2}, with the exception of the saGSpD operator pool that is not universal, all ADAPT-VQE simulations yielded numerically exact results for \ce{CH2}/STO-6G.
As might have been anticipated, the simulations employing the fully symmetry-adapted pDint0 operator pool exhibited the best performance.
Not only were all symmetries rigorously enforced, but also both chemically accurate and numerically exact energetics were obtained with the least number of parameters, consistent with the dimensions of the pertinent fully symmetry-adapted Hilbert spaces [152 $A_1$ ($C_\text{2v}$) CSFs for $\theta_\text{HCH} = 60^\circ$ and 93 $A_\text{g}$ ($D_\text{2h}$) CSFs for $\theta_\text{HCH} = 180^\circ$].

The ADAPT-VQE-GSD simulation distinguishes the $\tensor*[^{1}]{A}{_{1}}$ and $\tensor*[^{3}]{B}{_{1}}$ states by enforcing point-group symmetry.
Nevertheless, the convergence to the target state is slower than the pDint0 case, as the simulation explores a larger Hilbert space spanned by 321 $A_1$ ($C_\text{2v}$) $S_z = 0$ Slater determinants for $\theta_\text{HCH} = 60^\circ$ and 169 $A_g$ ($D_\text{2h}$) $S_z = 0$ Slater determinants for $\theta_\text{HCH} = 180^\circ$.
It is worth mentioning that the spin contamination is not as severe as in the \ce{H6} case.
This can be attributed to the fact that the lowest-energy totally symmetric non-singlet state is higher in energy by \SI{465}{\millihartree} ($\tensor*[^{3}]{A}{_{1}}$) and \SI{683}{\millihartree} ($\tensor*[^{3}]{A}{_{g}}$) for $\theta_\text{HCH} = 60^\circ$ and $180^\circ$, respectively.

In contrast, ADAPT-VQE-saGSpD-full successfully targets the $\tensor*[^{1}]{A}{_{1}}$ state by enforcing $S^2$ symmetry.
Since spatial symmetry is not conserved, the explored Hilbert space is spanned by 490 singlet CSFs.
Similar to the GSD case, the symmetry contamination is minor.
In the $\theta_\text{HCH} = 60^\circ$ geometry, this behavior can be explained by the large energy gap of \SI{156}{\millihartree} between the target $\tensor*[^{1}]{A}{_{1}}$ state and the lowest-energy spatial-symmetry contaminant, $\tensor*[^{1}]{A}{_{2}}$.
For the linear conformation where we employed $D_\text{2h}$ symmetry, the situation is more complicated as the target $\tensor*[^{1}]{A}{_{\text{g}}}$ state is degenerate with the lowest-energy $\tensor*[^{1}]{B}{_{2\text{g}}}$ state.
In light of this, one would naturally anticipate the ADAPT-VQE-saGSpD-full wavefunction to gain a substantial $\tensor*[^{1}]{B}{_{2\text{g}}}$ character.
However, in a minimum STO-6G basis set description, there exists only one $B_\text{2g}$ single excitation and thus it is not included in our extended set of pairs and triplets of spatial-symmetry-violating singles.
The next viable symmetry contaminants ($\tensor*[^{1}]{B}{_{1\text{u}}}$, $\tensor*[^{1}]{B}{_{3\text{u}}}$) are \SI{276}{\millihartree} higher in energy than $\tensor*[^{1}]{A}{_{\text{g}}}$, rationalizing the negligible spatial-symmetry contamination.

\subsection{Scenario III: Crossing of States with a Single Symmetry Difference}
\label{subsec:BeH2_BO}

Finally, we examine the worst-case scenario, in which the two crossing states belong to different irreducible representations of only one symmetry.
The first example in this category is the $C_\text{s}$-symmetric bending in a distorted \ce{BeH2} molecule.
The geometry was obtained as follows.
We first optimized the geometry of the lowest-energy, totally symmetric singlet state of $D_{\infty\text{h}}$-symmetric \ce{BeH2} at the FCI/STO-6G level of theory, obtaining $R_\text{Be--H} = \SI{1.310011}{\angs}$.
Subsequently, we elongated one of the two Be--H bonds by \SI{0.733008}{\angs}, the FCI/STO-6G equilibrium bond length of \ce{H2}.
The H--Be--H angle was then varied from $0^\circ$ to $180^\circ$.
As shown in panel (c) of \cref{fig:fci}, the two lowest-energy states are of $\tensor*[^{1}]{A}{^{\prime}}$ and $\tensor*[^{3}]{A}{^{\prime}}$ symmetries and cross twice.
As a result, $\tensor*[^{3}]{A}{^{\prime}}$ is the ground state for $\sim 38^\circ \le \theta_\text{HBeH} \le \sim 72^\circ$, switching to $\tensor*[^{1}]{A}{^{\prime}}$ for all other angles.
At $\theta_\text{HBeH} = 0^\circ$, the $\tensor*[^{1}]{A}{^{\prime}}$ and $\tensor*[^{3}]{A}{^{\prime}}$ states correlate with $\tensor*[^{1}]{\Sigma}{^{+}}$ ($A_1$ in $C_\text{2v}$) and $\tensor*[^{3}]{\Pi}{}$ ($B_1$ component in $C_\text{2v}$), respectively.
For $\theta_\text{HBeH} = 180^\circ$, the $\tensor*[^{1}]{A}{^{\prime}}$ and $\tensor*[^{3}]{A}{^{\prime}}$ states correlate with $\tensor*[^{1}]{\Sigma}{^{+}}$ ($A_1$ in $C_\text{2v}$) and $\tensor*[^{3}]{\Sigma}{^{+}}$ ($A_1$ in $C_\text{2v}$), respectively.
To target the $\tensor*[^{1}]{A}{^{\prime}}$ state even when $\tensor*[^{3}]{A}{^{\prime}}$ lies lower in energy, the quantum algorithm must enforce $S^2$ symmetry.
We performed ADAPT-VQE simulations for three representative geometries that encompass all relevant regions of the PEC, namely, $\theta_\text{HBeH} = 0^\circ$, $50^\circ$, and $180^\circ$.
These angles were selected because the singlet--triplet gap reaches its maximum value in the [$0^\circ$, $38^\circ$], [$38^\circ$, $72^\circ$], and [$72^\circ$, $180^\circ$] regions.

\begin{figure*}[htpb]
    \centering
    \includegraphics[width=6.5in]{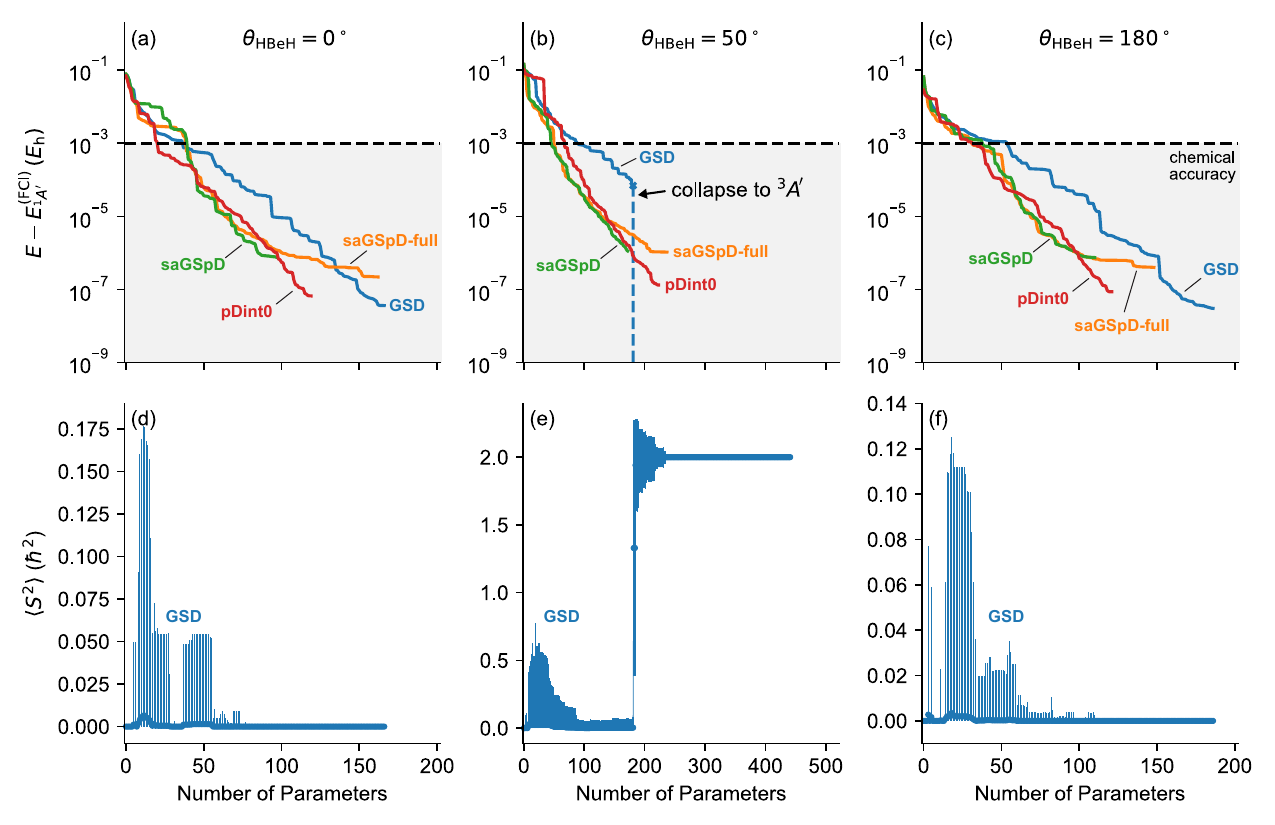}
    \caption{Convergence of ADAPT-VQE/STO-6G simulations using the GSD, saGSpD-full, saGSpD, and pDint0 operator pools for the \ce{BeH2} molecule at the $\theta_{\text{HBeH}}= 0^\circ$, $50^\circ$, and $180^\circ$ geometries, shown as functions of the number of parameters:
(a--c) Energy errors relative to the FCI/STO-6G energy of the lowest-energy $\tensor*[^{1}]{A}{^{\prime}}$ state ($\tensor*[^{1}]{A}{_{1}}$ for the linear geometries) and
(d--f) expectation values of the total spin squared operator and their standard deviations using the GSD pool.
Note that, in this case, the ADAPT-VQE-saGSpD-full simulations do not break spatial symmetry (see main text).
    }
    \label{fig:beh2_1}
\end{figure*}

As shown in \cref{fig:beh2_1}, the ADAPT-VQE simulations based on the universal operator pools pDint0 and saGSpD-full that enforce $S^2$ symmetry converged to the target $\tensor*[^{1}]{A}{^{\prime}}$ ($\tensor*[^{1}]{A}{_{1}}$ for linear conformations) state for all examined geometries. 
When employing the GSD operator pool, which, although universal, violates $S^2$, ADAPT-VQE was well-behaved in the geometries where the target state is the global energy minimum, eventually producing numerically exact results.
For the $\theta_\text{HBeH} = 50^\circ$ conformation, ADAPT-VQE-GSD was initially tracking the desired $\tensor*[^{1}]{A}{^{\prime}}$ state, but eventually collapsed to $\tensor*[^{3}]{A}{^{\prime}}$, which is lower in energy.
This is further corroborated by the dramatic increase in the expectation value of $S^2$, jumping from $\sim \SI{0}{\hbar^2}$, consistent with a singlet state, to $\sim \SI{2}{\hbar^2}$, correlating with triplet spin multiplicity.

Paradoxically, for this particular system, the ADAPT-VQE simulations paired with the fully symmetry-adapted, yet not universal, saGSpD operator pool produced numerically exact results for all examined geometries.
Nevertheless, this surprising result can be rationalized by examining the symmetry of the orbitals.
Starting with the $\theta_\text{HBeH} = 50^\circ$ geometry, all orbitals have $A^\prime$ symmetry, except one that has $A^{\prime\prime}$.
As the target state is totally symmetric, the eigenvector is a linear combination of CSFs in which the $A^{\prime\prime}$ spatial orbital is either unoccupied or doubly occupied.
This occupancy pattern can be fully described by the Lie algebra of spin-adapted singles and perfect-pairing doubles, explaining why saGSpD is exact in this case.
Similar reasoning applies when focusing on the linear conformations.
In these cases, all orbitals have $A_1$ symmetry, except a single $B_1$ and a single $B_2$ orbitals.
Considering that the direct product of $B_1$ and $B_2$ equals $A_2$, and that no orbital has $A_2$ symmetry, the totally symmetric Hilbert space is spanned by singlet CSFs in which each of the $B_1$ and $B_2$ spatial orbitals is either unoccupied or doubly occupied, similar to the $\theta_\text{HBeH} = 50^\circ$ geometry.
Therefore, the excellent performance of the fully symmetry-adapted ADAPT-VQE-saGSpD approach for this system is fortuitous and a direct consequence of the adopted minimum-basis-set description.
This also explains the absence of spatial-symmetry breaking in the ADAPT-VQE-saGSpD-full simulations for this system. 

The next molecular process we examine is the dissociation of the BO diatomic.
As shown in panel (d) of \cref{fig:fci}, the two lowest-energy FCI/STO-6G states have $\tensor*[^{2}]{\Sigma}{^{+}}$ and $\tensor*[^{2}]{\Pi}{}$ symmetries and cross around $R_\text{B--O} = \SI{1.6}{\angs}$.
Consequently, $\tensor*[^{2}]{\Sigma}{^{+}}$ is the ground electronic state for $R_\text{B--O} \le \SI{1.6}{\angs}$, with $\tensor*[^{2}]{\Pi}{}$ being lower in energy as the dissociation limit is approached.
To target the $\tensor*[^{2}]{\Sigma}{^{+}}$ state even when $\tensor*[^{2}]{\Pi}{}$ lies lower in energy, the quantum algorithm must enforce point-group symmetry.
We performed ADAPT-VQE simulations at two representative points of the PEC.
The first one is $R_\text{B--O} = \SI{1.2}{\angs}$, which is around the equilibrium region where $\tensor*[^{2}]{\Sigma}{^{+}}$ is the ground state.
The second one is at $R_\text{B--O} = \SI{2.1}{\angs}$, which is in the recoupling region where $\tensor*[^{2}]{\Pi}{}$ is lower in energy.

\begin{figure*}[htpb]
    \centering
        \includegraphics[width=5in]{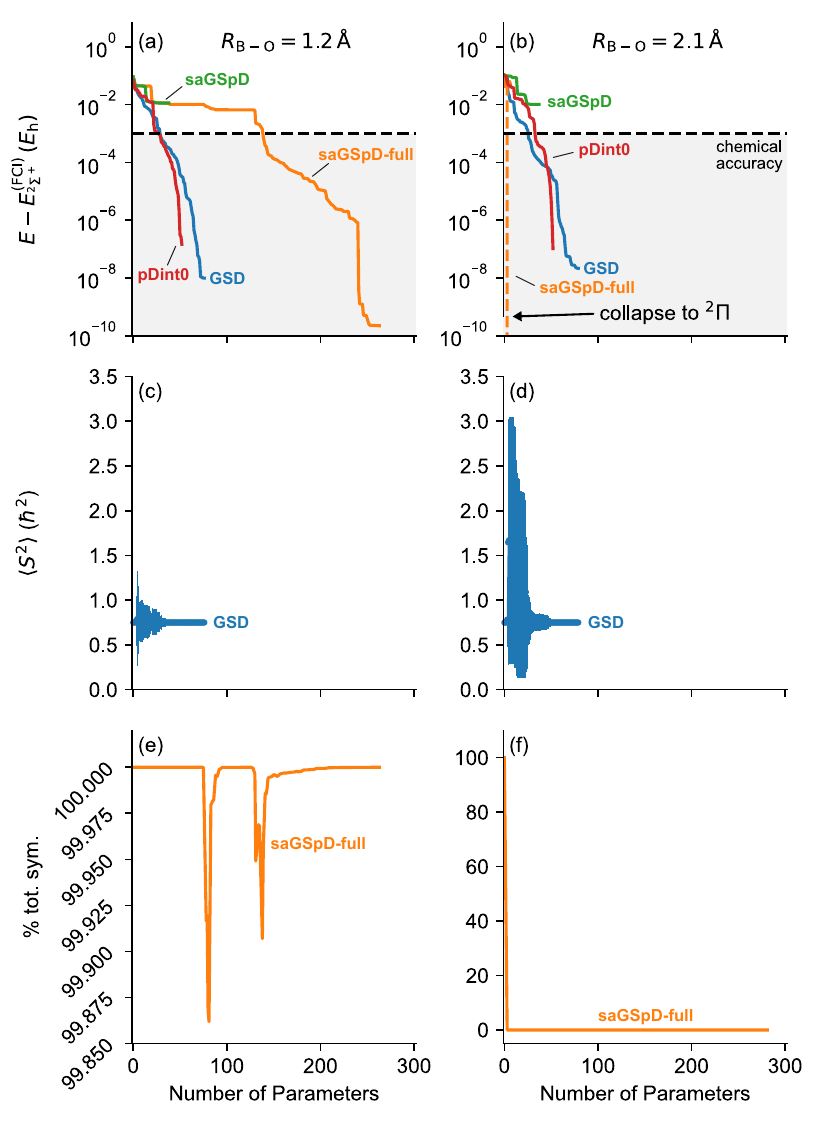}
    \caption{Convergence of ADAPT-VQE/STO-6G simulations using the GSD, saGSpD-full, saGSpD, and pDint0 operator pools for the \ce{BO} molecule at the $R_{\text{B--O}}= \SI{1.2}{\angs}$ and \SI{2.1}{\angs} geometries, shown as functions of the number of parameters:
(a,b) Energy errors relative to the FCI/STO-6G energy of the lowest-energy $\tensor*[^{2}]{\Sigma}{^{+}}$ state;
(c,d) expectation values of the total spin squared operator and their standard deviations using the GSD pool; and
(e,f) weight of the totally symmetric component using the saGSpD-full pool.
    }
    \label{fig:bo_1}
\end{figure*}

As might have been anticipated, the ADAPT-VQE simulation based on the non-universal saGSpD approach failed to provide chemically accurate results for any of the examined geometries (see \cref{fig:bo_1}).
Both ADAPT-VQE simulations paired with the universal pDint0 and GSD operator pools that respect spatial symmetry yielded numerically accurate energies for both internuclear distances.
The situation is more complicated when we turn to the results obtained with the saGSpD-full pool, which violates spatial symmetry.
For the $R_\text{B--O} = \SI{1.2}{\angs}$ internuclear distance, where the target $\tensor*[^{2}]{\Sigma}{^{+}}$ state is lower in energy, the ADAPT-VQE-saGSpD-full simulation produces numerically accurate energies, albeit with a slower convergence than when the pDint0 and GSD operator pools are employed.
However, the picture changes dramatically when the $R_\text{B--O} = \SI{2.1}{\angs}$ geometry is considered.
Here, ADAPT-VQE-saGSpD-full immediately collapses to the $\tensor*[^{2}]{\Pi}{}$ state (a mixture of $B_1$ and $B_2$ components in $C_\text{2v}$), which is lower in energy.

\section{Conclusions}\label{sec:conc}

In this work, we examined the interplay between universality, symmetry adaptation, and circuit efficiency in quantum simulations of molecular systems.
To emphasize the trade-off between these three desirable qualities, we focused on three types of operator pools.
The first one involved universal fully symmetry-adapted operator pools that lead to deep and long quantum circuits.
The second type contained operator pools that are gate-friendly and fully symmetry-adapted but not universal.
The third category was concerned with operator pools that are universal and gate-friendly at the expense of symmetry breaking.

On the theoretical front, we showed that enforcing point-group symmetry in the gate-efficient saGSpD operator pool compromises its universality.
In particular, we showed that for standard Abelian molecular point groups, classes of singlet spin-adapted doubles lie outside the Lie algebra generated by spin-adapted singles and perfect-pairing doubles.
The universality can be restored by incorporating spatial-symmetry-violating spin-adapted singles into the pool, albeit risking symmetry contamination and potential collapse to a state with undesired symmetries.

 To determine when symmetry-breaking operator pools can be safely used and when strict symmetry enforcement is mandatory, we examined the cases where the target state is either the global energy minimum or it crosses with another state with which they differ in one or multiple symmetries.
Our formal and numerical results lead to several practical guidelines for the design and use of operator pools in quantum simulations for chemistry.
First, when the target state is the global ground state, universal pools that break symmetries can still be employed, but fully symmetry-adapted pools offer faster and more predictable convergence.
Second, when multiple symmetries distinguish the crossing states, enforcing at least one of them is sufficient to avoid variational collapse.
Third, when crossing states differ in only a single symmetry property, that symmetry must be enforced to prevent symmetry contamination and potential variational collapse.
Finally, while non-universal, fully symmetry-adapted pools, such as saGSpD, can sometimes perform surprisingly well, their success is not guaranteed and may rely on accidental features of the orbital and symmetry structure.

In this work, we focused on imposing symmetries at the level of the unitaries defining the quantum circuits.
In the future, we will explore the usefulness of combining unitaries that are partially symmetry-adapted with alternative approaches to numerically enforce the broken symmetries.
Characteristic examples of such strategies include projection techniques,\cite{Izmaylov.2019.10.1021/acs.jpca.9b01103} particularly those of the variation-after-projection type, and Hamiltonian penalty terms.\cite{Kuroiwa.2021.10.1103/PhysRevResearch.3.013197}

\section*{Supplementary Material}

The {\sm} document contains a discussion of the Lie algebra generated by the elements of the saGSpD pool, and figures displaying additional ADAPT-VQE numerical results for the \ce{H6}/STO-6G system.

\section*{Acknowledgments}

This work was supported by the U.S.\ Department of Energy under Award No.\ DE-SC0019374.

\clearpage

\renewcommand{\theequation}{S\arabic{equation}}
\setcounter{equation}{0}

\renewcommand{\thetable}{S\arabic{table}}
\setcounter{table}{0}

\renewcommand{\thefigure}{S\arabic{figure}}
\setcounter{figure}{0}

\renewcommand{\thesection}{S\arabic{section}}
\setcounter{section}{0}

\renewcommand{\thepage}{S\arabic{page}}
\setcounter{page}{1}

\onecolumngrid
\fontsize{12}{22}\selectfont
\begin{center}
	\textbf{\large Supplemental Material:\\
		Symmetry Dilemmas in Quantum Computing for Chemistry:\linebreak A Comprehensive Analysis
	}\\[.2cm]
	Ilias Magoulas$^{*}$, Muhan Zhang, and Francesco A.\ Evangelista\\[.1cm]
	{\itshape Department of Chemistry and Cherry Emerson Center for Scientific Computation,\\ 
		Emory University, Atlanta, Georgia 30322, USA\\}
	${}^*$Corresponding author; e-mail: ilias.magoulas@emory.edu.
\end{center}

\newpage

The supplementary material is organized as follows.
In \cref{ssec:lie}, we discuss the Lie algebra generated by the elements of the singlet spin-adapted singles and perfect-pairing doubles (saGSpD) operator pool.
In \cref{ssec:h6}, we report additional numerical results for the \ce{H6}/STO-6G linear chain obtained with ADAPT-VQE using the saGSpD and generalized singles and doubles (GSD) operator pools.

\pagebreak

\section{Lie Algebra of the Singlet Spin-Adapted Singles and Perfect-Pairing Doubles Operator Pool}\label{ssec:lie}

To illustrate how the Lie algebra of the saGSpD pool contains elements of the parent singlet spin-adapted GSD pool, we provide the pertinent expressions up to triply nested commutators.
In what follows, we assume that the spatial orbital indices are distinct and ordered alphabetically, namely, $P<Q<R<S<T$.
Expressions marked in blue font color indicate spin-adapted double excitations outside the saGSpD operator pool.

We begin with the case of a single commutator.
The commutator of two spin-adapted singles yields, at most, another spin-adapted single,
\begin{align}
	\comm{A_P^Q}{A_R^S} &= 0,\\
	\comm{A_P^Q}{A_Q^R} &= -\frac{1}{\sqrt{2}} A_P^R,\\
	\comm{A_P^Q}{A_P^Q} &= 0.
\end{align}
Consequently, the singlet spin-adapted singles form a Lie subalgebra of the full Lie algebra generated by saGSpD operators.
Depending on the number of common indices, a single commutator of two perfect-pairing doubles results in
\begin{align}
	\comm{A_{PP}^{QQ}}{A_{RR}^{SS}} &= 0,\\
	\begin{split}\label{seq:cond_pp}
		\comm{A_{PP}^{QQ}}{A_{QQ}^{RR}} &= (n_{\Qu} + n_{\Qd} - 1) A_{PP}^{RR}\\
		&= \tensor*[^{[0,\frac{1}{2}]}]{A}{_{PPQ}^{RRQ}} - A_{PP}^{RR},
	\end{split}\\
	\comm{A_{PP}^{QQ}}{A_{PP}^{QQ}} &= 0.
\end{align}
In the case of a single common index [\cref{seq:cond_pp}], the commutator produces a conditional perfect-pairing double that is nonzero only when the spectator spatial orbital $Q$ is not singly occupied.
Alternatively, this can also be viewed as a linear combination of a perfect-pairing double and a singlet spin-adapted triple excitation involving a spectator single excitation.
The last type of single commutator is between a perfect-pairing double and a spin-adapted single,
\begin{align}
	\comm{A_{PP}^{QQ}}{A_R^S} &= 0\\
	\textcolor{blue}{\comm{A_{PP}^{QQ}}{A_Q^R}} &= -A_{PP}^{QR} \label{seq:ppqr}\\
	\begin{split}\label{seq:cond_s}
		\textcolor{blue}{\comm{A_{PP}^{QQ}}{A_P^Q}} &= (n_{\Pd} - n_{\Qd}) A_{\Pu}^{\Qu} + (n_{\Pu} - n_{\Qu}) A_{\Pd}^{\Qd}\\
		&= A_{PP}^{PQ} + A_{QQ}^{PQ}.
	\end{split}
\end{align}
For two common indices [\cref{seq:cond_s}], the commutator yields a conditional single excitation that is equivalent to a linear combination of two spin-adapted double excitations.
From the above considerations, we see that excitations of the form $A_{PP}^{QR}$ can be obtained from single commutators between a perfect-pairing double and a spin-adapted single.
However, note that spin-adapted double excitations with three common indices are linearly dependent [see \cref{seq:cond_s}].

Having exhausted all single commutators, we next consider doubly nested commutators.
The innermost commutators are those that generate operators outside the saGSpD operator pool, namely, \cref{seq:cond_s,seq:ppqr,seq:cond_pp}.
We obtain:
\begin{align}
	\textcolor{blue}{\comm{\comm{A_{PP}^{QQ}}{A_{QQ}^{RR}}}{A_P^S}} & = (n_{\Qu} + n_{\Qd} - 1) A_{RR}^{PS}\\
	\comm{\comm{A_{PP}^{QQ}}{A_{QQ}^{RR}}}{A_Q^S} & = -A_{PP}^{RR} H_Q^S \\
	\textcolor{blue}{\comm{\comm{A_{PP}^{QQ}}{A_{QQ}^{RR}}}{A_R^S}} & = (1 - n_{\Qu} - n_{\Qd}) A_{PP}^{RS}\\
	\textcolor{blue}{\comm{\comm{A_{PP}^{QQ}}{A_{QQ}^{RR}}}{A_P^Q}} & = \frac{1}{\sqrt{2}}(n_{\Pd} + n_{\Qu} - 1) A_{\Ru\Rd}^{\Pu\Qd} - \frac{1}{\sqrt{2}}(n_{\Pu} + n_{\Qd} - 1) A_{\Ru\Rd}^{\Pd\Qu}\\
	\textcolor{blue}{\comm{\comm{A_{PP}^{QQ}}{A_{QQ}^{RR}}}{A_P^R}} & = -\frac{1}{\sqrt{2}} (1 - n_{\Qu} - n_{\Qd}) \left[ A_{\Pu}^{\Qu} (n_{\Pd} - n_{\Rd}) + A_{\Pd}^{\Qd} (n_{\Pu} - n_{\Ru}) \right]\\
	\textcolor{blue}{\comm{\comm{A_{PP}^{QQ}}{A_{QQ}^{RR}}}{A_Q^R}} & = (n_{\Qd} + n_{\Ru} - 1) A_{\Pu\Pd}^{\Qu\Rd} - (n_{\Qu} + n_{\Rd} - 1) A_{\Pu\Pd}^{\Qd\Ru} \\
	\comm{\comm{A_{PP}^{QQ}}{A_{QQ}^{RR}}}{A_{PP}^{SS}} & = -(1 - n_{\Pu} - n_{\Pd}) (1 - n_{\Qu} - n_{\Qd}) A_{RR}^{SS}\\
	\comm{\comm{A_{PP}^{QQ}}{A_{QQ}^{RR}}}{A_{QQ}^{SS}} & = -2A_{PP}^{RR} H_{QQ}^{SS}\\
	\comm{\comm{A_{PP}^{QQ}}{A_{QQ}^{RR}}}{A_{RR}^{SS}} & = (1 - n_{\Qu} - n_{\Qd}) (1 - n_{\Ru} - n_{\Rd}) A_{PP}^{SS}\\
	\comm{\comm{A_{PP}^{QQ}}{A_{QQ}^{RR}}}{A_{PP}^{QQ}} & = (1 - n_{\Pu} - n_{\Pd} + 2n_{\Pu}n_{\Pd}) A_{QQ}^{RR} \\
	\comm{\comm{A_{PP}^{QQ}}{A_{QQ}^{RR}}}{A_{PP}^{RR}} & = 0\\
	\comm{\comm{A_{PP}^{QQ}}{A_{QQ}^{RR}}}{A_{QQ}^{RR}} & = -(1 - n_{\Ru} - n_{\Rd} + 2n_{\Ru}n_{\Rd}) A_{PP}^{QQ}\\
	\textcolor{blue}{\comm{\comm{A_{PP}^{QQ}}{A_Q^R}}{A_P^S}} & = \tensor*[^{[0]}]{A}{_{PS}^{QR}}\\
	\textcolor{blue}{\comm{\comm{A_{PP}^{QQ}}{A_Q^R}}{A_Q^S}} & = \frac{1}{\sqrt{2}} A_{PP}^{RS}\\
	\textcolor{blue}{\comm{\comm{A_{PP}^{QQ}}{A_Q^R}}{A_R^S}} & = \frac{1}{\sqrt{2}} A_{PP}^{QS}\\
	\textcolor{blue}{\comm{\comm{A_{PP}^{QQ}}{A_Q^R}}{A_P^Q}} & = -\frac{1}{\sqrt{2}} A_{PP}^{PR} + \tensor*[^{[0]}]{A}{_{PQ}^{QR}} \\
	\textcolor{blue}{\comm{\comm{A_{PP}^{QQ}}{A_Q^R}}{A_P^R}} & = -\frac{1}{\sqrt{2}} A_{PP}^{PQ} + \tensor*[^{[0]}]{A}{_{PR}^{QR}}\\
	\comm{\comm{A_{PP}^{QQ}}{A_Q^R}}{A_Q^R} & = A_{PP}^{RR} - A_{PP}^{QQ}\\
	\textcolor{blue}{\comm{\comm{A_{PP}^{QQ}}{A_Q^R}}{A_{PP}^{SS}}} & = (1 - n_{\Pu} - n_{\Pd}) A_{SS}^{QR} \\
	\comm{\comm{A_{PP}^{QQ}}{A_Q^R}}{A_{QQ}^{SS}} & = - \tensor*[^{[0,\frac{1}{2}]}]{A}{_{PPQ}^{SSR}}\\
	\comm{\comm{A_{PP}^{QQ}}{A_Q^R}}{A_{RR}^{SS}} & = - \tensor*[^{[0,\frac{1}{2}]}]{A}{_{PPR}^{SSQ}}\\
	\textcolor{blue}{\comm{\comm{A_{PP}^{QQ}}{A_Q^R}}{A_{PP}^{QQ}}} & = n_{\Pu\Pd} A_Q^R + \frac{1}{\sqrt{2}}(1 - n_{\Pu} - n_{\Pd}) \left( n_{\Qd} A_{\Qu}^{\Ru} + n_{\Qu} A_{\Qd}^{\Rd} \right)\\
	\textcolor{blue}{\comm{\comm{A_{PP}^{QQ}}{A_Q^R}}{A_{PP}^{RR}}} & = -n_{\Pu\Pd} A_Q^R - \frac{1}{\sqrt{2}}(1 - n_{\Pu} - n_{\Pd}) \left( n_{\Rd} A_{\Qu}^{\Ru} + n_{\Ru} A_{\Qd}^{\Rd} \right)\\
	\comm{\comm{A_{PP}^{QQ}}{A_Q^R}}{A_{QQ}^{RR}} & = 0\\
	\textcolor{blue}{\comm{\comm{A_{PP}^{QQ}}{A_P^Q}}{A_P^R}} & = \frac{1}{2} \left[ (n_{\Pd} - n_{\Qd}) A_{\Qu}^{\Ru} + (n_{\Pu} - n_{\Qu}) A_{\Qd}^{\Rd} - \sqrt{2} A_{PP}^{QR} - A_{\Pu\Qd}^{\Pd\Ru} - A_{\Pd\Qu}^{\Pu\Rd}\right]\\
	\textcolor{blue}{\comm{\comm{A_{PP}^{QQ}}{A_P^Q}}{A_Q^R}} & = \frac{1}{2} \left[ (n_{\Qd} - n_{\Pd}) A_{\Pu}^{\Ru} + (n_{\Qu} - n_{\Pu}) A_{\Pd}^{\Rd} - \sqrt{2} A_{QQ}^{PR} + A_{\Pu\Qd}^{\Qu\Rd} + A_{\Pd\Qu}^{\Qd\Ru}\right]\\
	\comm{\comm{A_{PP}^{QQ}}{A_P^Q}}{A_P^Q} & = -2 A_{PP}^{QQ}\\
	\textcolor{blue}{\comm{\comm{A_{PP}^{QQ}}{A_P^Q}}{A_{PP}^{RR}}} & = \frac{1}{\sqrt{2}} \left[ (1 - n_{\Pu} - n_{\Qd}) A_{\Ru\Rd}^{\Pd\Qu} - (1 - n_{\Pd} - n_{\Qu}) A_{\Ru\Rd}^{\Pu\Qd} \right]\\
	\textcolor{blue}{\comm{\comm{A_{PP}^{QQ}}{A_P^Q}}{A_{QQ}^{RR}}} & = \frac{1}{\sqrt{2}} \left[ (1 - n_{\Pu} - n_{\Qd}) A_{\Ru\Rd}^{\Pd\Qu} - (1 - n_{\Pd} - n_{\Qu}) A_{\Ru\Rd}^{\Pu\Qd} \right]\\
	\textcolor{blue}{\comm{\comm{A_{PP}^{QQ}}{A_P^Q}}{A_{PP}^{QQ}}} & = (n_{\Pd} + n_{\Qd} - 2 n_{\Pd}n_{\Qd}) A_{\Pu}^{\Qu} + (n_{\Pu} + n_{\Qu} - 2n_{\Pu}n_{\Qu}) A_{\Pd}^{\Qd}
\end{align}
It is worth mentioning that spin-adapted double excitations of the form $\tensor*[^{[0]}]{A}{_{PQ}^{RS}}$ already appear in singly nested commutators.

To arrive at spin-adapted double excitations going through an intermediate triplet, we need to consider multiply nested commutators of saGSpD elements.
In what follows, we give the expressions of those triply nested commutators that generate spin-polarized double excitations, \textit{i.e.}, excitations of the form $A_{\Pu\Qu}^{\Ru\Su}$ and $A_{\Pd\Qd}^{\Rd\Sd}$.
The corresponding expressions are:
\begin{align}
	\begin{split}\label{seq:triple1}
		\textcolor{blue}{\comm{\comm{\comm{A_{PP}^{QQ}}{A_{QQ}^{RR}}}{A_Q^S}}{A_P^R}} &= \frac{1}{2} \left[ (n_{\Pd} - n_{\Rd}) \left( -A_{\Pu\Qu}^{\Ru\Su} - A_{\Pu\Qd}^{\Ru\Sd} + A_{\Pu\Su}^{\Qu\Ru} + A_{\Pu\Sd}^{\Qd\Ru} \right)\right.\\
		&\phantom{={}\frac{1}{2}}\left.+ (n_{\Pu} - n_{\Ru}) \left( -A_{\Pd\Qd}^{\Rd\Sd} - A_{\Pd\Qu}^{\Rd\Su} + A_{\Pd\Sd}^{\Qd\Rd} + A_{\Pd\Su}^{\Qu\Rd}\right)\right]
	\end{split}\\
	\begin{split}\label{seq:triple2}
		\textcolor{blue}{\comm{\comm{\comm{A_{PP}^{QQ}}{A_{QQ}^{RR}}}{A_P^R}}{A_Q^S}} &= \frac{1}{2} \left[ (n_{\Pd} - n_{\Rd}) \left( -A_{\Pu\Qu}^{\Ru\Su} - A_{\Pu\Qd}^{\Ru\Sd} + A_{\Pu\Su}^{\Qu\Ru} + A_{\Pu\Sd}^{\Qd\Ru} \right)\right.\\
		&\phantom{={}\frac{1}{2}}\left.+ (n_{\Pu} - n_{\Ru}) \left( -A_{\Pd\Qd}^{\Rd\Sd} - A_{\Pd\Qu}^{\Rd\Su} + A_{\Pd\Sd}^{\Qd\Rd} + A_{\Pd\Su}^{\Qu\Rd}\right)\right]
	\end{split}\\
	\begin{split}\label{seq:triple3}
		\textcolor{blue}{\comm{\comm{\comm{A_{PP}^{QQ}}{A_{Q}^{R}}}{A_P^S}}{A_{PP}^{QQ}}} &= \frac{1}{2} \left[ A_{\Pu\Ru}^{\Qu\Su}(n_{\Qd} - n_{\Pd}) + A_{\Pu\Rd}^{\Qd\Su} (n_{\Qu} - n_{\Pd}) \right.\\
		&\phantom{={}\frac{1}{2}}\left. + A_{\Pd\Rd}^{\Qd\Sd} (n_{\Qu} - n_{\Pu}) + A_{\Pd\Ru}^{\Qu\Sd} (n_{\Qd} - n_{\Pu}) \right]
	\end{split}\\
	\begin{split}\label{seq:triple4}
		\textcolor{blue}{\comm{\comm{\comm{A_{PP}^{QQ}}{A_{Q}^{R}}}{A_P^S}}{A_{PP}^{RR}}} &= \frac{1}{2} \left[ A_{\Pu\Qu}^{\Ru\Su}(n_{\Rd} - n_{\Pd}) + A_{\Pu\Qd}^{\Rd\Su} (n_{\Ru} - n_{\Pd}) \right.\\
		&\phantom{={}\frac{1}{2}}\left. + A_{\Pd\Qd}^{\Rd\Sd} (n_{\Ru} - n_{\Pu}) + A_{\Pd\Qu}^{\Ru\Sd} (n_{\Rd} - n_{\Pu}) \right]
	\end{split}\\
	\begin{split}\label{seq:triple5}
		\textcolor{blue}{\comm{\comm{\comm{A_{PP}^{QQ}}{A_{Q}^{R}}}{A_P^S}}{A_{QQ}^{SS}}} &= \frac{1}{2} \left[ A_{\Pu\Qu}^{\Ru\Su}(n_{\Qd} - n_{\Sd}) + A_{\Pu\Qd}^{\Rd\Su} (n_{\Qu} - n_{\Sd}) \right.\\
		&\phantom{={}\frac{1}{2}}\left. + A_{\Pd\Qd}^{\Rd\Sd} (n_{\Qu} - n_{\Su}) + A_{\Pd\Qu}^{\Ru\Sd} (n_{\Qd} - n_{\Su}) \right]
	\end{split}\\
	\begin{split}\label{seq:triple6}
		\textcolor{blue}{\comm{\comm{\comm{A_{PP}^{QQ}}{A_{Q}^{R}}}{A_P^S}}{A_{RR}^{SS}}} &= \frac{1}{2} \left[ A_{\Pu\Ru}^{\Qu\Su}(n_{\Rd} - n_{\Sd}) + A_{\Pu\Rd}^{\Qd\Su} (n_{\Ru} - n_{\Sd}) \right.\\
		&\phantom{={}\frac{1}{2}}\left. + A_{\Pd\Rd}^{\Qd\Sd} (n_{\Ru} - n_{\Su}) + A_{\Pd\Ru}^{\Qu\Sd} (n_{\Rd} - n_{\Su}) \right]
	\end{split}\\
	\begin{split}\label{seq:triple7}
		\textcolor{blue}{\comm{\comm{\comm{A_{PP}^{QQ}}{A_{Q}^{R}}}{A_{QQ}^{SS}}}{A_P^S}} &= \frac{1}{2} \left[ (n_{\Pd} - n_{\Sd}) \left( A_{\Pu\Qu}^{\Ru\Su} + A_{\Pu\Qd}^{\Rd\Su}\right)\right.\\
		&\phantom{={}\frac{1}{2} }\left.+ (n_{\Pu} - n_{\Su}) \left( A_{\Pd\Qd}^{\Rd\Sd} + A_{\Pd\Qu}^{\Ru\Sd}\right)\right]
	\end{split}\\
	\begin{split}\label{seq:triple8}
		\textcolor{blue}{\comm{\comm{\comm{A_{PP}^{QQ}}{A_{Q}^{R}}}{A_{RR}^{SS}}}{A_P^S}} &= \frac{1}{2} \left[ (n_{\Pd} - n_{\Sd}) \left( A_{\Pu\Ru}^{\Qu\Su} + A_{\Pu\Rd}^{\Qd\Su}\right)\right.\\
		&\phantom{={}\frac{1}{2} }\left.+ (n_{\Pu} - n_{\Su}) \left( A_{\Pd\Rd}^{\Qd\Sd} + A_{\Pd\Ru}^{\Qu\Sd}\right)\right]
	\end{split}\\
	\begin{split}\label{seq:triple9}
		\textcolor{blue}{\comm{\comm{\comm{A_{PP}^{QQ}}{A_{Q}^{R}}}{A_{PP}^{QQ}}}{A_{P}^{S}}} &= \frac{1}{2} \left[ (n_{\Pd} - n_{\Qd}) \left(A_{\Pu\Qu}^{\Ru\Su} - A_{\Pu\Ru}^{\Qu\Su}\right) +
		(n_{\Pd} - n_{\Qu}) \left(A_{\Pu\Qd}^{\Rd\Su} - A_{\Pu\Rd}^{\Qd\Su}\right) \right.\\
		&\phantom{={}\frac{1}{2}}\left. + (n_{\Pu} - n_{\Qu}) \left(A_{\Pd\Qd}^{\Rd\Sd} - A_{\Pd\Rd}^{\Qd\Sd}\right) +
		(n_{\Pu} - n_{\Qd}) \left(A_{\Pd\Qu}^{\Ru\Sd} - A_{\Pd\Ru}^{\Qu\Sd}\right) \right]
	\end{split}\\
	\begin{split}\label{seq:triple10}
		\textcolor{blue}{\comm{\comm{\comm{A_{PP}^{QQ}}{A_{Q}^{R}}}{A_{PP}^{RR}}}{A_{P}^{S}}} &= \frac{1}{2} \left[ (n_{\Pd} - n_{\Rd}) \left(- A_{\Pu\Qu}^{\Ru\Su} + A_{\Pu\Ru}^{\Qu\Su}\right) +
		(n_{\Pd} - n_{\Ru}) \left(- A_{\Pu\Qd}^{\Rd\Su} + A_{\Pu\Rd}^{\Qd\Su}\right) \right.\\
		&\phantom{={}\frac{1}{2}}\left. + (n_{\Pu} - n_{\Ru}) \left(- A_{\Pd\Qd}^{\Rd\Sd} + A_{\Pd\Rd}^{\Qd\Sd}\right) +
		(n_{\Pu} - n_{\Rd}) \left(- A_{\Pd\Qu}^{\Ru\Sd} + A_{\Pd\Ru}^{\Qu\Sd}\right) \right]
	\end{split}
\end{align}
As shown in the main text, conditional excitations of the form \cref{seq:triple1,seq:triple2,seq:triple3,seq:triple4,seq:triple5,seq:triple6,seq:triple7,seq:triple8,seq:triple9,seq:triple10} give rise to linear combinations of the $\tensor*[^{[0]}]{A}{_{PQ}^{RS}}$ and $\tensor*[^{[1]}]{A}{_{PQ}^{RS}}$ double excitations.
All of these triply nested commutators involve two spin-adapted singles and two perfect-pairing doubles.
If the saGSpD pool respects spatial symmetry, \textit{i.e.}, its elements belong to the totally symmetric irreducible representation of the point group, double excitations of the form $\tensor*[^{[1]}]{A}{_{PQ}^{RS}}$ in which spatial orbitals $P$ and $Q$ share the same irreducible representation (and similarly for $R$ and $S$) are absent from the Lie algebra of the fully symmetry-adapted saGSpD operator pool.
\pagebreak

\section{Additional Numerical Results for $\mathbf{H}_\mathbf{6}$/STO-6G}\label{ssec:h6}
\subsection{ADAPT-VQE-GSD}
\FloatBarrier

\begin{figure*}[h!]
	\centering
	\includegraphics[width=3.2in]{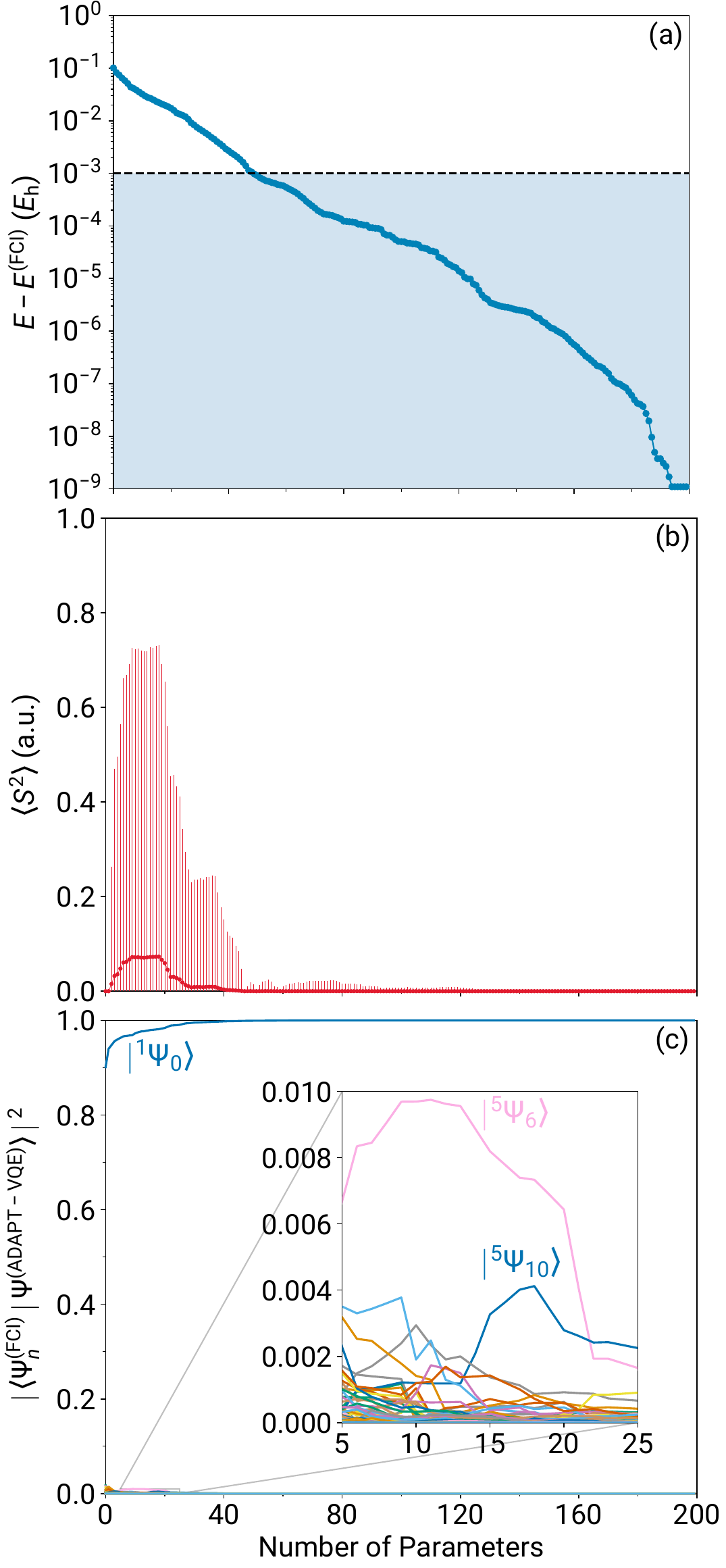}
	\caption{
		Convergence of ADAPT-VQE-GSD/STO-6G simulations for the \ce{H6} linear chain at the $R_\text{H--H} = \SI{1.0}{\text{\AA}}$ geometry, shown as functions of the number of parameters:
		(a) Energy errors relative to FCI/STO-6G; (b) expectation values of the total spin squared operator and their standard deviations; and (c) weights of the totally symmetric, $S_z = 0$ FCI eigenvectors.
	}
\end{figure*}

\begin{figure}[h]
	\centering
	\includegraphics[width=3.33in]{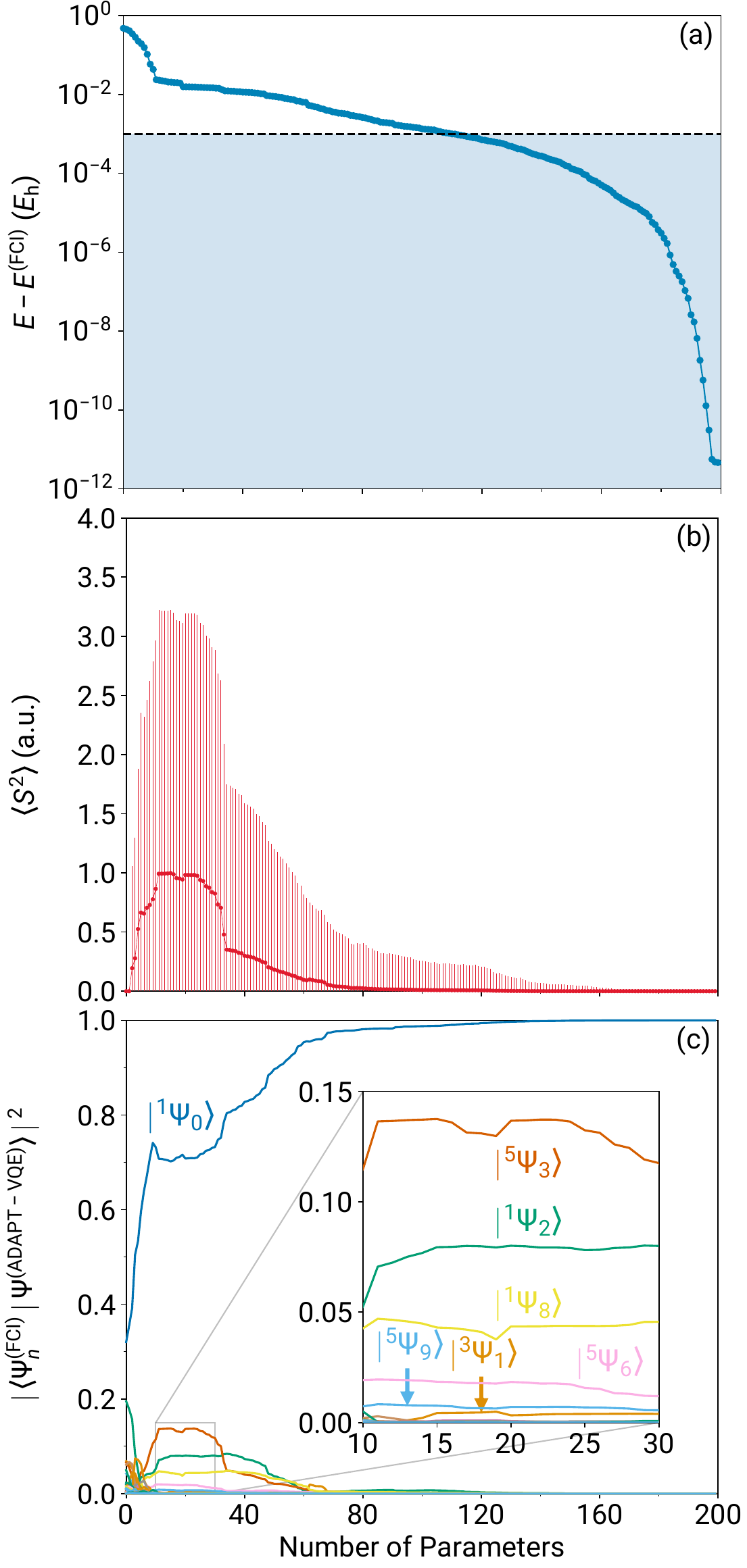}
	\caption{
		Convergence of ADAPT-VQE-GSD/STO-6G simulations for the \ce{H6} linear chain at the $R_\text{H--H} = \SI{2.0}{\text{\AA}}$ geometry, shown as functions of the number of parameters:
		(a) Energy errors relative to FCI/STO-6G; (b) expectation values of the total spin squared operator and their standard deviations; and (c) weights of the totally symmetric, $S_z = 0$ FCI eigenvectors.
	}
\end{figure}

\begin{figure*}[h!]
	\centering
	\includegraphics[width=3.33in]{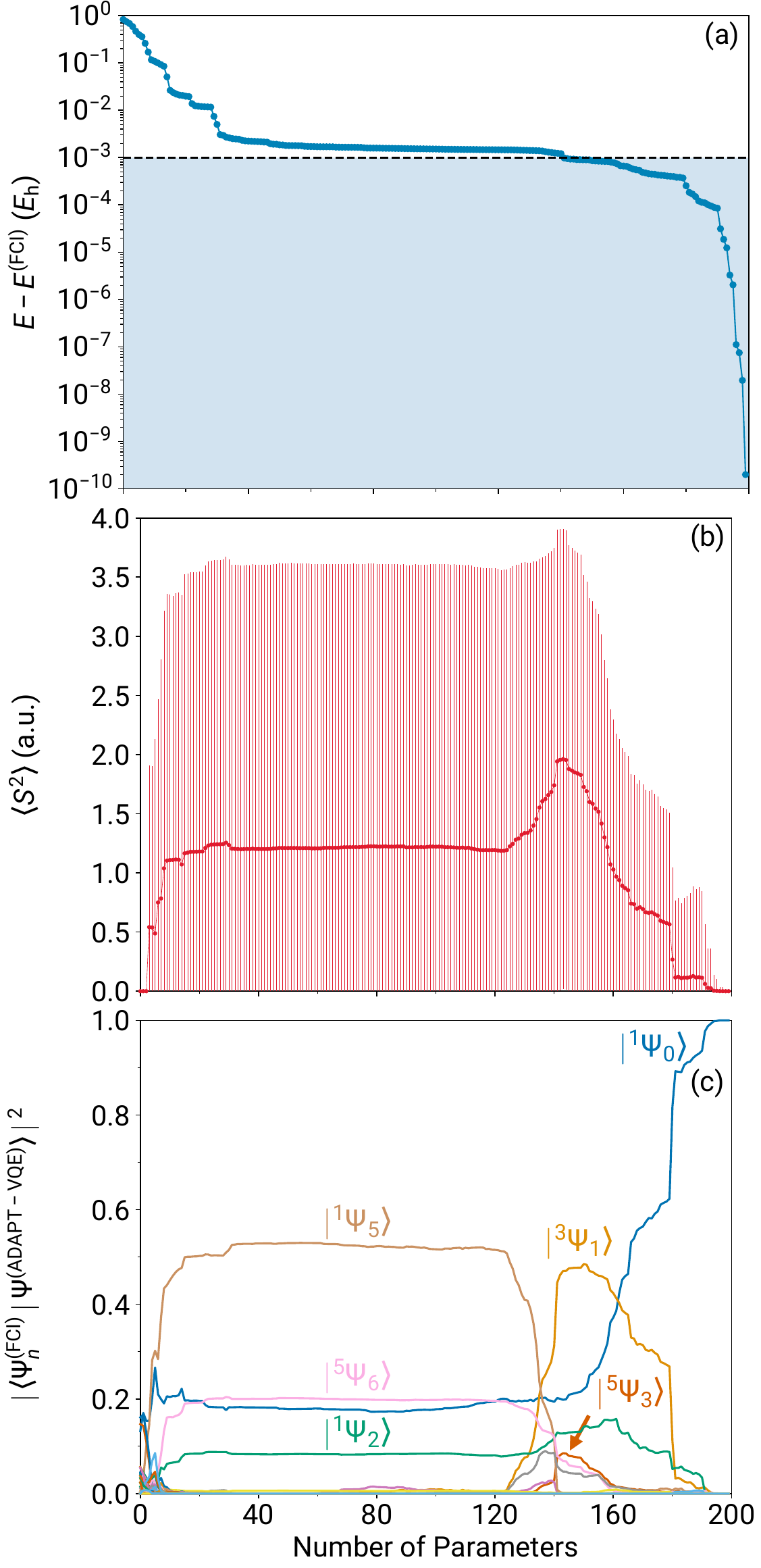}
	\caption{
		Convergence of ADAPT-VQE-GSD/STO-6G simulations for the \ce{H6} linear chain at the $R_\text{H--H} = \SI{3.0}{\text{\AA}}$ geometry, shown as functions of the number of parameters:
		(a) Energy errors relative to FCI/STO-6G; (b) expectation values of the total spin squared operator and their standard deviations; and (c) weights of the totally symmetric, $S_z = 0$ FCI eigenvectors.
	}
\end{figure*}
\FloatBarrier

\subsection{ADAPT-VQE-saGSpD}
\FloatBarrier

\begin{figure*}[h!]
	\centering
	\includegraphics[width=3.33in]{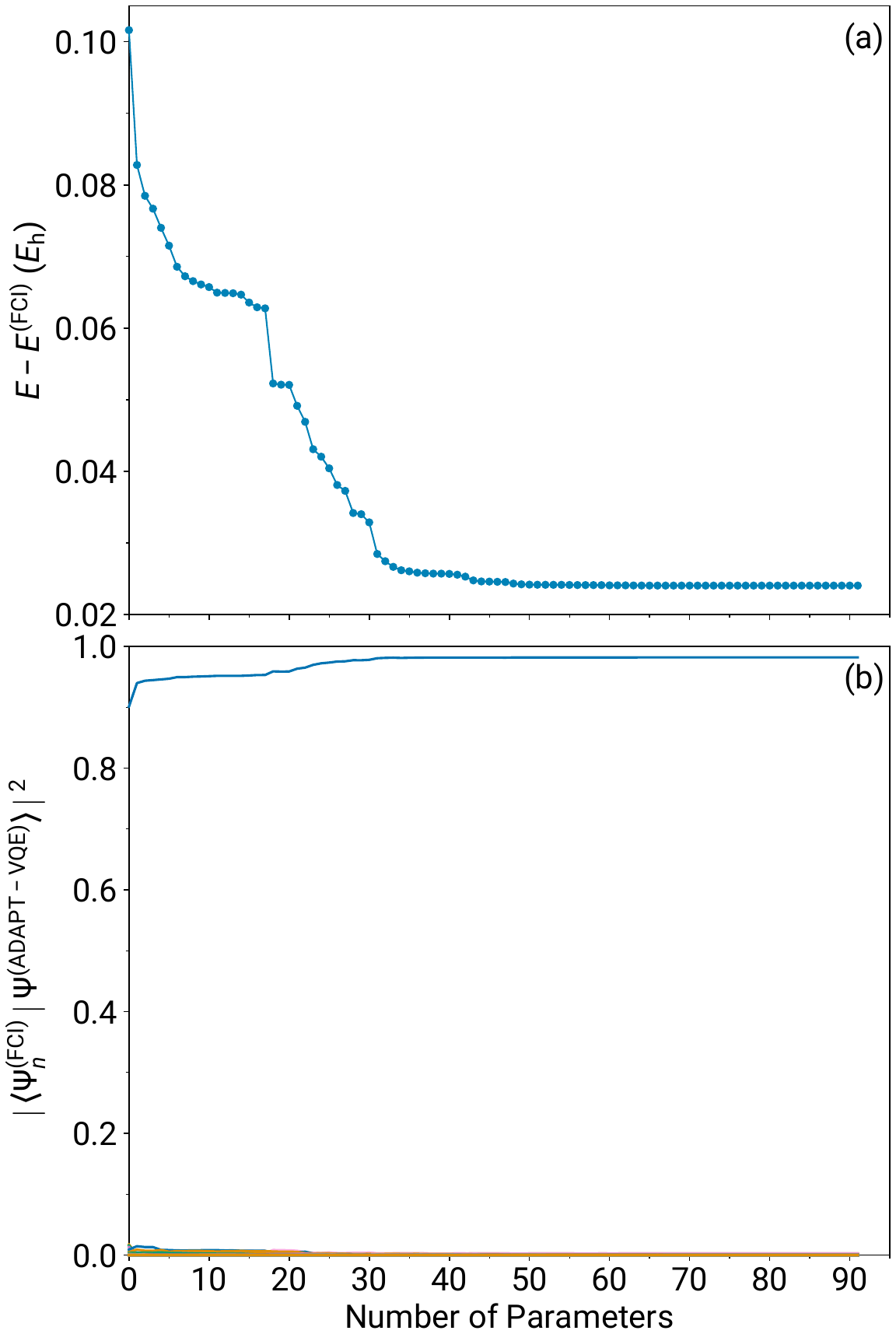}
	\caption{
		Convergence of fully symmetry-adapted ADAPT-VQE-saGSpD/STO-6G simulations for the \ce{H6} linear chain at the $R_\text{H--H} = \SI{1.0}{\text{\AA}}$ geometry, shown as functions of the number of parameters:
		(a) Energy errors relative to FCI/STO-6G and (b) weights of the totally symmetric, singlet FCI eigenvectors.
	}
\end{figure*}

\begin{figure*}[h!]
	\centering
	\includegraphics[width=\textwidth]{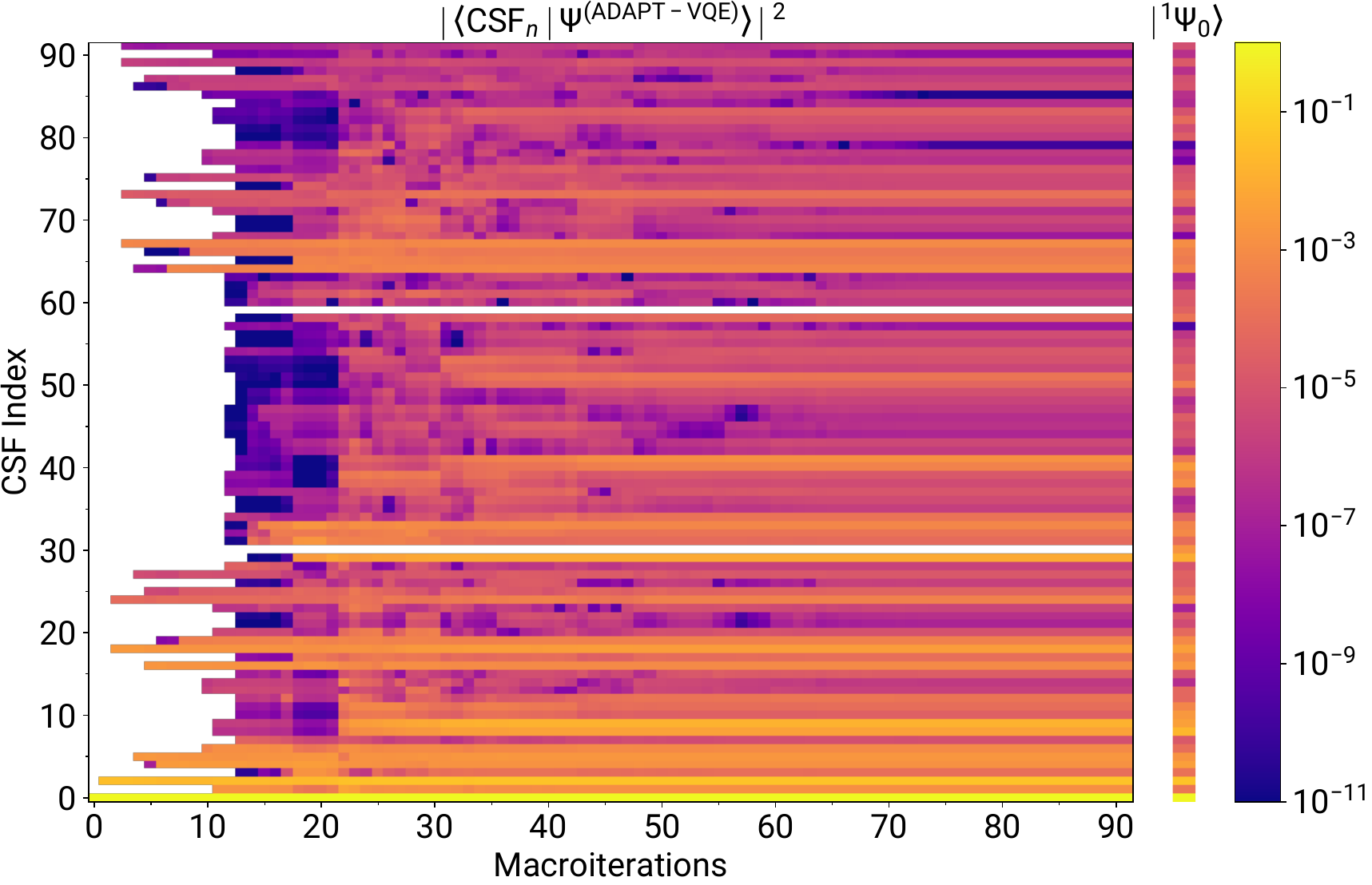}
	\caption{
		Comparison of CSF weights in the fully symmetry-adapted ADAPT-VQE-saGSpD wavefunction and the exact ground-state eigenvector for the $\text{H}_6$/STO-6G linear chain with $R_\text{H--H} = \SI{1.0}{\text{\AA}}$. The left heatmap shows CSF weights in the ADAPT-VQE-saGSpD wavefunction across macroiterations, with color intensity indicating weight magnitude on a shared logarithmic color scale. The right heatmap shows the corresponding CSF weights in the exact ground-state eigenvector.
	}
\end{figure*}

\begin{figure}[h]
	\centering
	\includegraphics[width=3.33in]{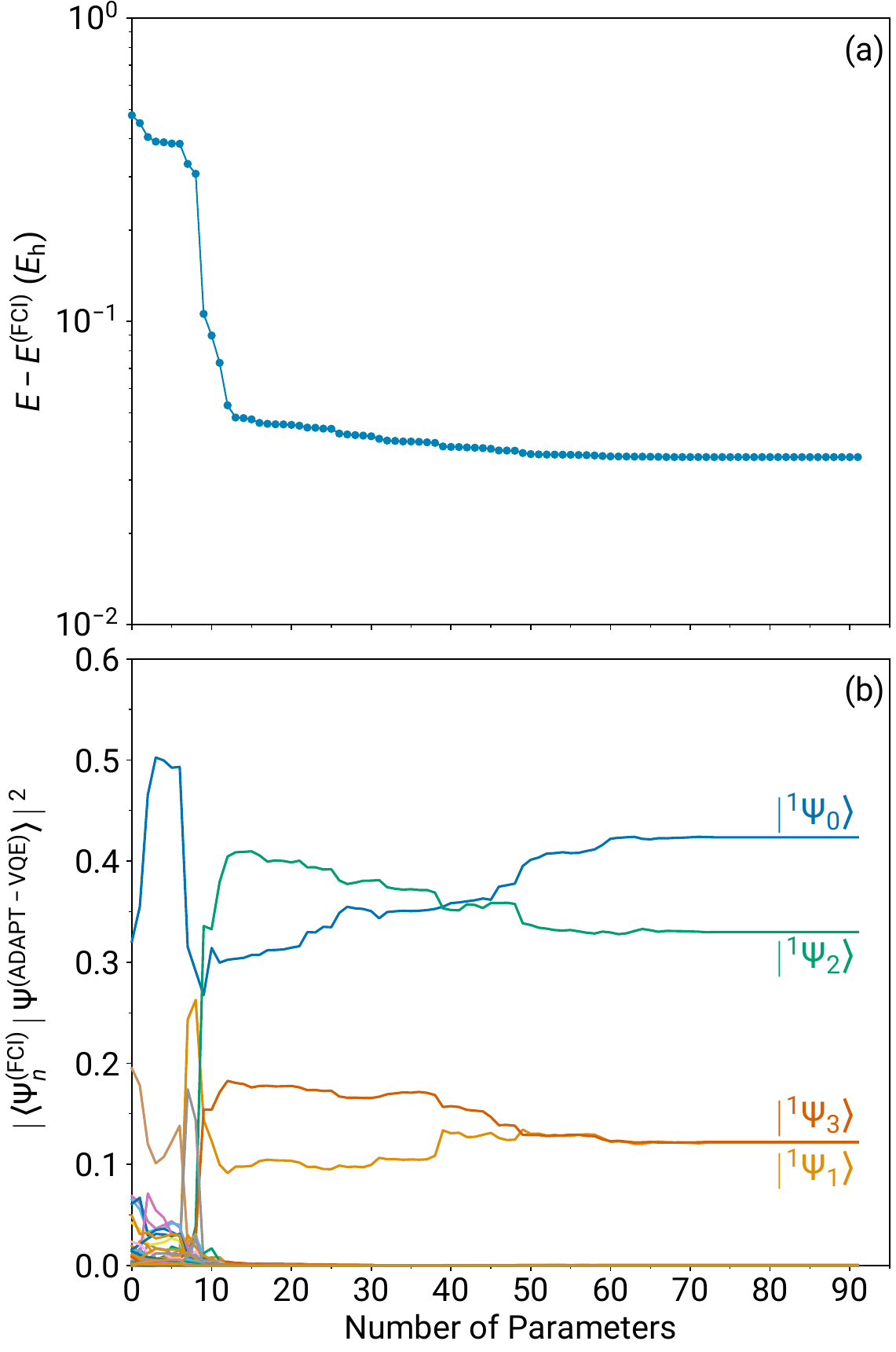}
	\caption{Convergence of fully symmetry-adapted ADAPT-VQE-saGSpD/STO-6G simulations for the \ce{H6} linear chain at the $R_\text{H--H} = \SI{2.0}{\text{\AA}}$ geometry, shown as functions of the number of parameters:
		(a) Energy errors relative to FCI/STO-6G and (b) weights of the totally symmetric, singlet FCI eigenvectors.}
	\label{fig:adapt-sagspd}
\end{figure}

\begin{figure*}[h!]
	\centering
	\includegraphics[width=\textwidth]{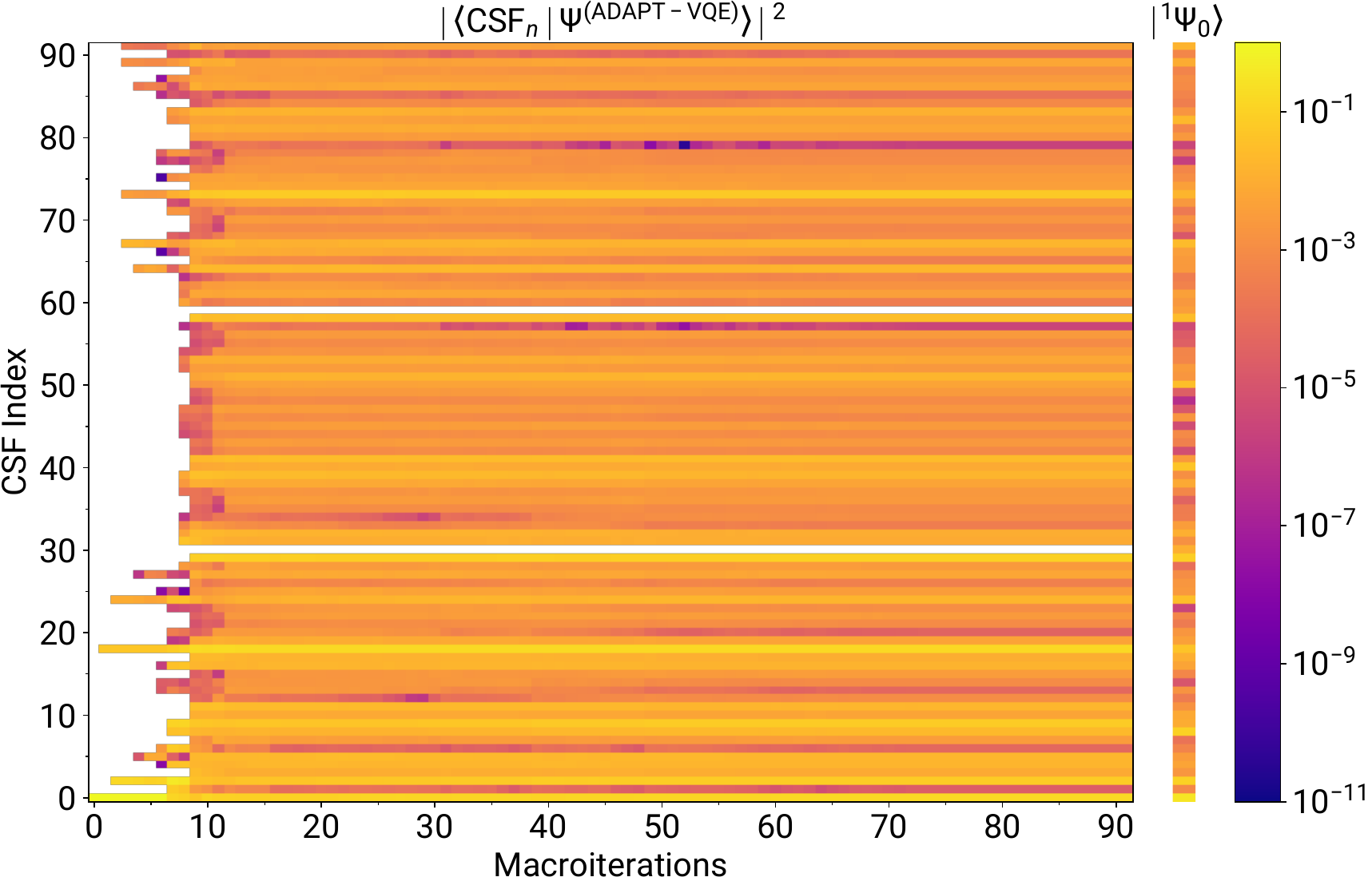}
	\caption{
		Comparison of CSF weights in the fully symmetry-adapted ADAPT-VQE-saGSpD wavefunction and the exact ground-state eigenvector for the $\text{H}_6$/STO-6G linear chain with $R_\text{H--H} = \SI{2.0}{\text{\AA}}$. The left heatmap shows CSF weights in the ADAPT-VQE-saGSpD wavefunction across macroiterations, with color intensity indicating weight magnitude on a shared logarithmic color scale. The right heatmap shows the corresponding CSF weights in the exact ground-state eigenvector.
	}
\end{figure*}

\begin{figure*}[h!]
	\centering
	\includegraphics[width=3.33in]{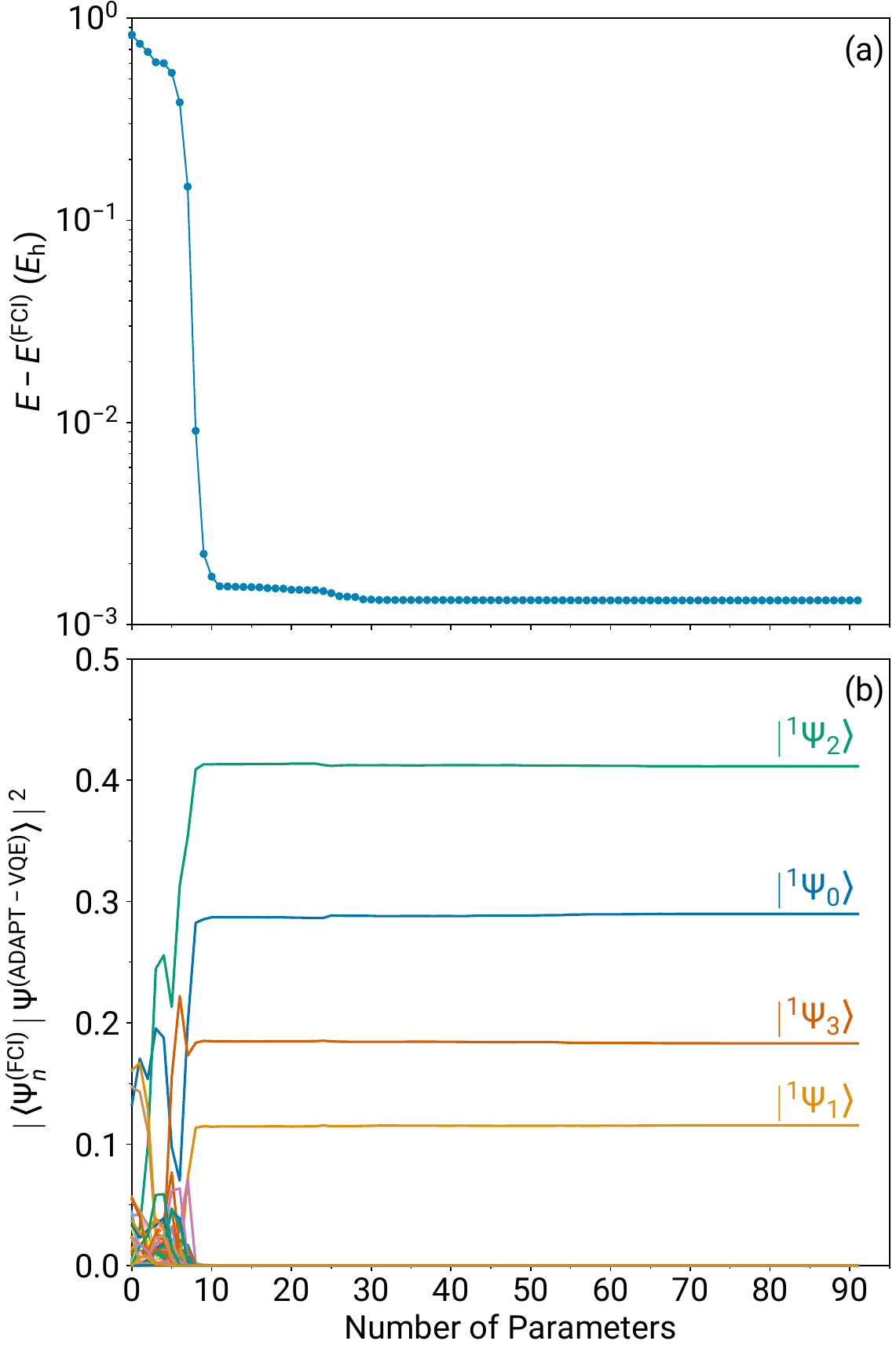}
	\caption{
		Convergence of fully symmetry-adapted ADAPT-VQE-saGSpD/STO-6G simulations for the \ce{H6} linear chain at the $R_\text{H--H} = \SI{3.0}{\text{\AA}}$ geometry, shown as functions of the number of parameters:
		(a) Energy errors relative to FCI/STO-6G and (b) weights of the totally symmetric, singlet FCI eigenvectors.
	}
\end{figure*}

\begin{figure*}[h!]
	\centering
	\includegraphics[width=\textwidth]{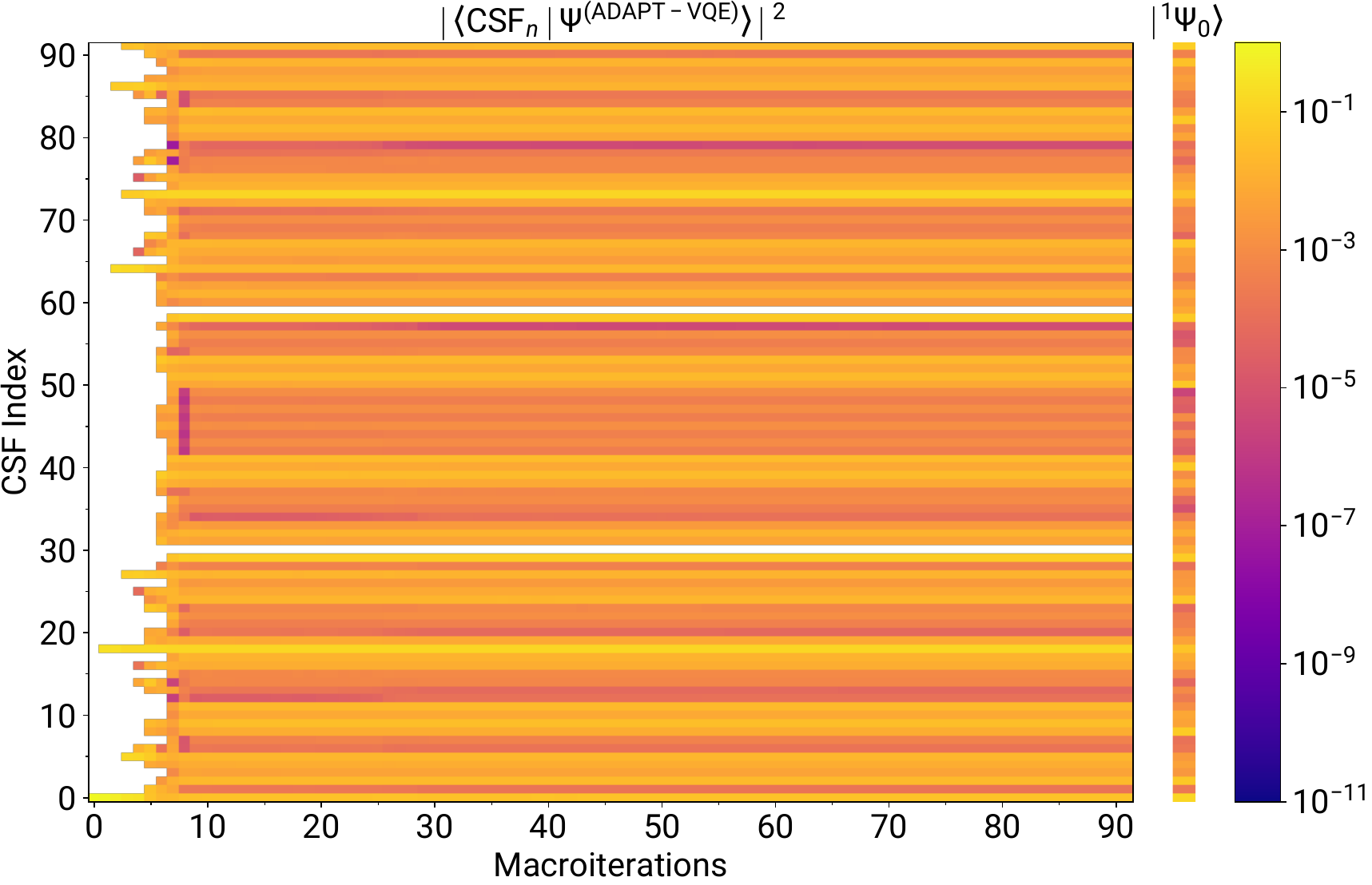}
	\caption{
		Comparison of CSF weights in the fully symmetry-adapted ADAPT-VQE-saGSpD wavefunction and the exact ground-state eigenvector for the $\text{H}_6$/STO-6G linear chain with $R_\text{H--H} = \SI{3.0}{\text{\AA}}$. The left heatmap shows CSF weights in the ADAPT-VQE-saGSpD wavefunction across macroiterations, with color intensity indicating weight magnitude on a shared logarithmic color scale. The right heatmap shows the corresponding CSF weights in the exact ground-state eigenvector.
	}
\end{figure*}

\end{document}